\documentclass[aps,prd,preprint,superscriptaddress,tightenlines,nofootinbib]{revtex4}

\long\def\symbolfootnote[#1]#2{\begingroup%
\def\thefootnote{\fnsymbol{footnote}}\footnote[#1]{#2}\endgroup}

\usepackage{amsfonts,amsmath,amsthm,amssymb}
\usepackage{mathrsfs}
\usepackage{bm}
\usepackage{color}
\usepackage{epsfig}

\newcommand{\PRE}[1]{{#1}}   % Use if preprint style

\newcommand{\beq}{\begin{equation}}
\newcommand{\eeq}{\end{equation}}
\newcommand{\bea}{\begin{flushleft} \begin{eqnarray}}
\newcommand{\eea}{\end{eqnarray}\end{flushleft}}

\newcommand{\postscript}[2]{\setlength{\epsfxsize}{#2\hsize}
   \centerline{\epsfbox{#1}}}
\newcommand{\comment}[1]{}

\newcommand{\er}[1]{\eqref{#1}}
\newcommand{\ci}[1]{}
\newcommand{\lsb}{\left[}
\newcommand{\rsb}{\right]}

\newcommand{\nn}{\nonumber \\}

\newcommand{\ba}{\begin{eqnarray}}
\newcommand{\ea}{\end{eqnarray}}
\newcommand{\be}{\begin{equation}}
\newcommand{\ee}{\end{equation}}
\newcommand{\bay}[1]{\left(\begin{array}{#1}}
\newcommand{\eay}{\end{array}\right)}

\def\xt{{\theta}}

\def\CD{{\cal D}}

\begin{document}

\preprint{
\hfil
\begin{minipage}[t]{3in}
\begin{flushright}
\vspace*{.4in}
MPP--2012--94\\
LMU-ASC 36/12\\
CERN-PH-TH/2012-162\\
\end{flushright}
\end{minipage}
}

\title{\PRE{\vspace*{0.55in}}
LHC Phenomenology  and Cosmology  of  String-Inspired Intersecting D-Brane Models}
%\PRE{\vspace*{0.1in}} }

\author{{\bf Luis A. Anchordoqui}}
\affiliation{Department of Physics,\\
University of Wisconsin-Milwaukee,
 Milwaukee, WI 53201, USA
\PRE{\vspace*{.05in}}
}

\author{{\bf Ignatios Antoniadis}}\thanks{On leave of absence
from CPHT Ecole Polytechnique, F-91128, Palaiseau Cedex.}
\affiliation{Department of Physics,\\ CERN Theory Division,
CH-1211 Geneva 23, Switzerland
\PRE{\vspace*{.05in}}}

\author{{\bf Haim \nolinebreak Goldberg}}
\affiliation{Department of Physics,\\
Northeastern University, Boston, MA 02115, USA
\PRE{\vspace*{.05in}}
}

\author{{\bf Xing Huang}}
\affiliation{School of Physics and Astronomy, \\
Seoul National University, Seoul 141-747, Korea
\PRE{\vspace*{.05in}}
}

\author{{\bf Dieter L\"ust}}
%\affiliation{Department of Physics,\\ CERN Theory Division,
%CH-1211 Geneva 23, Switzerland}

\affiliation{Max--Planck--Institut f\"ur Physik, \\ 
 Werner--Heisenberg--Institut,
80805 M\"unchen, Germany
\PRE{\vspace*{.05in}}
}

\affiliation{Arnold Sommerfeld Center for Theoretical Physics 
Ludwig-Maximilians-Universit\"at M\"unchen,
80333 M\"unchen, Germany
\PRE{\vspace{.05in}}
}

\author{{\bf Tomasz R. Taylor}}
\affiliation{Department of Physics,\\ CERN Theory Division,
CH-1211 Geneva 23, Switzerland
\PRE{\vspace*{.05in}}}

\affiliation{Department of Physics,\\
  Northeastern University, Boston, MA 02115, USA 
  \PRE{\vspace*{.05in}}
}

\author{{\bf Brian Vlcek}}
\affiliation{Department of Physics,\\
University of Wisconsin-Milwaukee,
 Milwaukee, WI 53201, USA
\PRE{\vspace*{.05in}}
}

%\PRE{\vspace*{.15in}}

\begin{abstract}\vskip 1mm
  \noindent We discuss the phenomenology and cosmology of a
  Standard-like Model inspired by string theory, in which the gauge
  fields are localized on D-branes wrapping certain compact cycles on
  an underlying geometry, whose intersection can give rise to chiral
  fermions. The energy scale associated with string physics is assumed
  to be near the Planck mass.  To develop our program in the simplest
  way, we work within the construct of a minimal model with
  gauge-extended sector $U (3)_B \times Sp (1)_L \times U (1)_{I_R}
  \times U (1)_L$.  The resulting $U (1)$ content gauges the baryon
  number $B$, the lepton number $L$, and a third additional abelian
  charge $I_R$ which acts as the third isospin component of an
  $SU(2)_R$.  All mixing angles and gauge couplings are fixed by
  rotation of the $U(1)$ gauge fields to a basis diagonal in
  hypercharge $Y$ and in an anomaly free linear combination of $I_R$
  and $B-L$.  The anomalous $Z'$ gauge boson obtains a string scale
  St\"uckelberg mass via a 4D version of the Green-Schwarz
  mechanism. To keep the realization of the Higgs mechanism minimal,
  we add an extra $SU(2)$ singlet complex scalar, which acquires a VEV
  and gives a TeV-scale mass to the non-anomalous gauge boson $Z''$.
  The model is fully predictive and can be confronted with dijet and
  dilepton data from LHC8 and, eventually, LHC14. We show that
  $M_{Z''} \approx 3 - 4~{\rm TeV}$ saturates current limits from the
  CMS and ATLAS collaborations. We also show that for $M_{Z''} \alt
  5~{\rm TeV}$, LHC14 will reach discovery sensitivity $\agt
  5\sigma$. After that, we demostrate in all generality that $Z''$
  milli-weak interactions could play an important role in
  observational cosmology. Finally, we examine some phenomenological
  aspects of the supersymmetric extension of the D-brane construct.

 \end{abstract}

\maketitle

\section{Introduction}

With the turn on of the Large Hadron Collider (LHC) at CERN, a new era
of discovery has just begun~\cite{ATLAS:2012ae,Chatrchyan:2012tx,ATLASnew,CMSnew}.
The $SU(3)_C \times SU(2)_L \times U(1)_Y$ Standard Model (SM) of
electroweak and strong interactions was once again severely tested
with a dataset corresponding to an integrated luminosity of $\sim
5~{\rm fb}^{-1}$ of $pp$ collisions collected at $\sqrt{s} = 8~{\rm
  TeV}$. The LHC8 data have shown no evidence for new physics beyond
the SM.  

However, there is another side to the story. The concordance model of
cosmology --the flat expanding universe containing 5\% baryons, 20\%
dark matter, and 75\% dark energy-- continues to be put on a firmer
footing through observations of the Supernova Search
Team~\cite{Riess:1998cb,Riess:2001gk,Tonry:2003zg}, the Supernova
Cosmology
Project~\cite{Perlmutter:1998np,Knop:2003iy,Kowalski:2008ez}, the
Wilkinson Microwave Anisotropy Probe
(WMAP)~\cite{Spergel:2003cb,Komatsu:2010fb}, the Sloan Digital Sky
Survey
(SDSS)~\cite{Abazajian:2003jy,Tegmark:2003ud,Abazajian:2008wr,Percival:2009xn},
and the Hubble Space Telescope~\cite{Riess:2009pu}. While not yet rock
solid experimentally, from these observations it is evident that in
order to describe the physics of the early universe, and thereupon
particle interactions at sub-fermi distances, new theoretical concepts
are necessary, which go beyond the SM.\footnote{It appears likewise
  from experimental evidence of neutrino flavor oscillations by the
  mixing of different mass eigenstates that the SM has to be
  extended~\cite{GonzalezGarcia:2007ib}.}

Arguably, another major driving force behind the consideration of physics
beyond the SM is the huge disparity between the strength of gravity
and of the SM forces. This hierarchy problem suggests that new physics
could be at play at the TeV-scale. To be more specific, the non-zero
vacuum expectation value of the scalar Higgs doublet  sets
the scale of electroweak interactions. However, due to the quadratic
sensitivity of the Higgs mass to quantum corrections from an
aribitrarily high mass scale, with no new physics between
the energy scale of electroweak unification, $M_{\rm EW} \sim 1~{\rm
  TeV}$, and the vicinity of the Planck mass, $M_{\rm Pl} \sim
10^{19}~{\rm GeV}$, the Higgs mass must be fine-tuned to an accuracy
of ${\cal O}(10^{32})$. Therefore, it is of interest to identify
univocal footprints that can plausible arise in theories with the
capacity to describe physics over this colossal range of scales.
Among various attempts in this direction, string theory is perhaps the
most successful candidate and also the most ambitious approach since
besides the SM gauge interactions it includes also the gravitational
force at the quantum level~\cite{Green:1987sp,Green:1987mn}.

In recent years there has been achieved substantial progress in
connecting string theory with particle physics and cosmology.
Important advances were fueled by the realization of the vital role
played by D-branes~\cite{Polchinski:1995mt,Polchinski:1996na} in connecting string theory to
phenomenology. This has
permitted the formulation of string theories with string scale
setting in at TeV scales, and together with large extra
dimensions~\cite{Antoniadis:1998ig}.

There are two peerless phenomenological consequences for TeV scale
D-brane string compatifications: the emergence of Regge recurrences at
parton collision energies $\sqrt{\hat s} \sim$ string scale $\equiv
M_s;$ and the presence of one or more additional $U(1)$ gauge
symmetries, beyond the $U(1)_Y$ of the SM. The latter follows from the
property that the gauge group for open strings terminating on a stack
of $N$ identical D-branes is $U(N)$ rather than $SU(N)$ for $N >
2$. (For $N = 2$ the gauge group can be $Sp(1) \cong SU(2)$ rather
than $U(2)$.)  In a series of recent publications we have exploited
both these properties to explore and anticipate new-physics signals
that could potentially be revealed at LHC. Regge recurrences most
distinctly manifest in the $\gamma +$
jet~\cite{Anchordoqui:2007da,Anchordoqui:2008ac} and
dijet~\cite{Lust:2008qc,Anchordoqui:2008di,Anchordoqui:2009mm,Dong:2010jt,Nath:2010zj}
spectra resulting from their decay.\footnote{The amplitudes of lowest
  massive Regge excitations that include $2 \to 2$ scattering
  processes involving 4 gauge bosons, or 2 gauge bosons and 2 quarks,
  are {\it universal}~\cite{Lust:2008qc}. Therefore, the $s$-channel
  pole terms of the average square amplitudes contributing to $\gamma
  +$ jet and dijet topologies can be obtained independent of the
  details of the compactification scheme.  For phenomenological
  purposes, the poles need to be softened to a Breit-Wigner form by
  obtaining and utilizing the correct total widths of the
  resonances~\cite{Anchordoqui:2008hi}. The recent search for such
  narrow resonances in data collected during the LHC8 run, 
  excludes a string scale below 4.69~TeV~\cite{cms_dijet8tev}.} The
extra $U(1)$ gauge symmetries beyond hypercharge have (in general)
triangle anomalies, but are cancelled by the Green-Schwarz
mechanism~\cite{Green:1984sg}. In addition there can be also massive
$U(1)$ gauge bosons, which are associated to 4D non-anomalous Abelian
gauge symmetries, but however originate from anomalous $U(1)$'s in six
dimensions.  In both cases, these $U(1)$ gauge bosons get
St\"uckelberg masses. Since in these D-brane models $M_s$ is assumed
to be ${\cal O} ({\rm TeV})$, the presence of these generic $U(1)$'s
may be amenable to experimental tests at the
LHC~\cite{Ghilencea:2002da,Berenstein:2008xg,Anchordoqui:2011eg}.

In this work we take a related but different approach studying new
physics effects of D-brane models with the conventional assumption
${\rm TeV} \ll M_s \alt M_{\rm Pl}$. The gauge symmetry also arises
from a product of $U(N)$ groups, guaranteeing extra $U(1)$ gauge
bosons in the spectrum. The weak hypercharge is identified with a
linear combination of anomalous $U(1)$'s which itself is anomaly free.
As indicated in the preceding paragraph, the extra anomalous $U(1)$
gauge bosons generically obtain a string scale St\"uckelberg mass. The
$U(1)$ symmetries survive as global selection rules in the effective
low energy theory. Such anomalous gauge bosons are now very heavy and
out of the LHC reach. 

However, in some D-brane models there exists
non-anomalous and also massless $U(1)$ gauge symmetries in addition to
hypercharge. Namely, under certain topological conditions the associated gauge bosons can
remain massless and obtain a low mass scale via the ordinary Higgs
mechanism. Some phenomenological aspects of these kind of $U(1)$ gauge bosons were recently 
discussed in~\cite{Cvetic:2011iq}.
In this paper we first revisit the prospects of detecting
such TeV-scale gauge bosons in particular at the LHC, and then we show in all
generality that their milli-weak interactions could play an important
role in observational cosmology. 

Before proceeding with an outline of the paper, we sketch some issues
surrounding the choice of a non-supersymmetric formulation. To avoid
the fine tuning inherent in the hierarchy problem, the overwhelmingly
favored approach is the introduction of supersymmetry (SUSY). However,
for the present study, this presents a difficult technical problem:
the full complexity of the scale of SUSY breaking has been pushed by
experiment into the TeV region, which coincides with the energy scale 
involved in searching for the extra $U(1)$ gauge bosons. In the
absence of an experimental signal for the onset of SUSY breaking, we
will extract from string theory the choice of the $U(1)$ gauge
assignments, as well as the quiver structure of the fermionic
couplings. In principle, SM-like non-SUSY vacua exist in the string
landscape~\cite{Bousso:2000xa,Susskind:2003kw,Douglas:2003um}. Throughout
most of this work we will operate within that vacuum structure.
However, before concluding we will also discuss in some detail the
phenomenology of supersymmetric vacua and the technical problems
associated with a phenomenologically viable breaking of an
additional $U(1)$ symmetry in a SUSY background.

The layout of the paper is as follows. In Sec.~\ref{SEC-II} we outline
the basic setting of intersecting D-brane models and discuss general
aspects of the effective low energy theory inhereted from properties
of the overarching string theory. After that, we particularize the
discussion to the $U(3)_B \times Sp(1)_L \times U(1)_L \times
U(1)_{I_R}$ intersecting D-brane configuration that realizes the SM by
open strings~\cite{Cremades:2003qj}. In Sec.~\ref{SEC-III} we study
the associated phenomenological aspects of non-anomalous $U(1)$ gauge
bosons related to experimental searches for new physics at the LHC.
In Sec.~\ref{SEC-IV} we explore cosmological predictions of
intersecting D-brane models in light of recent data, which seem to
favor the existence of roughly one additional neutrino species (in
addition to the 3 contained in the SM), challenging the earliest
observationally verified landmarks: big bang nucleosynthesis (BBN) and
the cosmic microwave background (CMB). The gist of Sec.~\ref{SEC-IV}
extends the previous study of TeV-scale string
compactifications~\cite{Anchordoqui:2011nh} to D-brane models where
some of the $U(1)$ masses are at a high string scale.  In
  Sec.~\ref{SEC-V} we examine the consequences of possible
  supersymmetric extensions. Our conclusions are collected in
Sec.~\ref{SEC-VI}.

\section{Standard Model from Intersecting D-branes}
\label{SEC-II}

D-brane string compactifications provide a collection of building
block rules that can be used to build up the SM or something very close
to
it~\cite{Aldazabal:2000sa,Blumenhagen:2000wh,Aldazabal:2000cn,Blumenhagen:2000ea,Ibanez:2001nd,Blumenhagen:2001te,Berenstein:2001nk,Cvetic:2001tj,Cvetic:2001nr,Antoniadis:2001np,Kiritsis:2002aj,Honecker:2004kb,Gmeiner:2005vz,Gmeiner:2008xq,Honecker:2012jd}. In
this section, we will briefly review the basics of constructing
such D-brane models. More comprehensive treatments can be found
in~\cite{Kiritsis:2003mc,Lust:2004ks,Blumenhagen:2005mu,Blumenhagen:2006ci}.

\subsection{Construction Rules and Generalities of D-brane Models}

The details of the D-brane construct depend a lot on whether we use
oriented string or unoriented string models. The basic unit of gauge
invariance for oriented string models is a $U(1)$ field, so that a
stack of $N$ identical D-branes eventually generates a $U(N)$ theory
with the associated $U(N)$ gauge group. In the presence of many
D-brane types, the gauge group becomes a product form $\prod U(N_P)$,
where $N_P$ reflects the number of D-branes in each stack. As an
illustration,  consider Type IIA string theory compactified on a six
dimensional manifold $\mathscr{M}$. A specific configuration will be
given by $K$ stacks of intersecting D6-branes filling 4-dimensional
Minkowski spacetime $M_4$ and wrapping internal homology 3-cycles of
$\mathscr{M}$. Each stack consists of $N_P$ coincident D6 branes whose
world-volume is $M_4 \times {\Pi}_P$, where ${\Pi}_P$ is the
corresponding homology class of each 3-cycle, with $P = 1, \dots, K$. The
closed string degrees of freedom reside in the entire ten dimensional
space, which in addition to the gravitational fields, contain the
geometric scalar moduli fields of the internal space. The open string
degrees of freedom give rise to the gauge theory on the D6-brane
world-volumes, with gauge group $\prod U(N_{P})$. In addition, there
are open string modes which split into states with both ends on the
same stack of branes as well as those connecting different stacks of
branes.  The latter are particularly interesting: there is a chiral
fermion living at each four-dimensional intersection of two branes $P$
and $Q$, transforming in the bifundamental representation of $U(N_P)
\times U(N_Q)$~\cite{Berkooz:1996km}. The intersection number of these
two branes, $I_{P Q} \equiv [{\Pi}_{P}] \cdot [{\Pi}_{Q}]$, is a
topologically invariant integer whose modulus gives us the
multiplicity of such massless fermionic content and its sign depends
on the chirality of such fermions. A particularly simple subfamily of
the configurations described above consist of taking $\mathscr{M}$ as
a factorizable six-torus: $T^6 = T^2 \times T^2 \times T^2$. We can
then further simplify the configurations assuming that the 3-cycles
can be factorized as three 1-cycles, each of them wrapping on a
different $T^2$. In this case the homology 3-cycle $\Pi_P$ can be
expressed as
\begin{equation}
[\Pi_P] = [(n_P^1, m_P^1), (n_P^2, m_P^2), (n_P^3, m_P^3)] \,,
\end{equation}
where $(n_P^i,m_P^i)$, are the wrapping numbers of each  
$D6_P$-brane, on the $i^{th}$ torus, with $n_P^i$ and ($m_P^i$) being the number 
of times the brane is wrapping around the $i^{th}$ torus. The intersection number takes a simple form
\begin{equation}
I_{PQ} = \prod_{i=1}^3 (n_P^i m_Q^i - m_P^i n_Q^i) \, .
\end{equation}

In orientifold brane configurations, which are necessary for tadpole
cancellation~\cite{Uranga:2000xp,Aldazabal:2000dg}, and thus
consistency of the theory, open strings become in general
non-oriented. For unoriented strings the above rules still apply, but
we are allowed many more choices because the branes come in two
different types. There are the branes whose images under the
orientifold are different from themselves and their image branes, and
also branes who are their own images under the orientifold
procedure. Stacks of the first type combine with their mirrors and
give rise to $U(N)$ gauge groups, while stacks of the second type give
rise to only $SO(N)$ or $Sp(N)$ gauge groups.

Generally speaking, intersecting D-brane models involve at least three
kinds of generic mass scales.  First, of course, there is the
fundamental string scale, 
\begin{equation}
M_s = \frac{1}{\sqrt{\alpha'}} \,,
\end{equation}
where $\alpha'$ is the slope parameter of the well known Regge trajectories of vibrating strings
\begin{equation}
j = j_0 + \alpha'  \, M^2 \,,
\end{equation}
with $j$ and $M = \sqrt{n} M_s$ the spin and mass of the resonant
state, respectively ($n = 1, \dots$). Second, compactification from ten to four
dimensions on an internal six--dimensional space of volume $V_6$
defines a mass scale:
\begin{equation}
M_6={1\over V_6^{1/6}}\, .
\end{equation}
Third, wrapping a stack $P$ of
D$(p+3)$-branes around the internal $p$-cycle defines
an internal world-volume $V_p^{(P)}=(2\pi)^p\ v_p^{(P)}$ of this D-branes stack and an associated (Kaluza-Klein)
mass:
\begin{equation}
M_p^{(P)}={1\over \left(v_p^{(P)} \right)^{1/p}}\, .
\end{equation}

These three types of fundamental dimensional parameters of D-brane
models are linked to four-dimensional physical observables. First, the
Planck mass given by
\begin{equation}
M_{\rm Pl}^2=8\ e^{-2\phi_{10}}\ M_s^8\ \frac{V_6}{(2\pi)^{6}}
\end{equation}
determines the strength of gravitational interactions. Here, the
dilaton field $\phi_{10}$ is related to the string coupling constant
through $g_s=e^{\phi_{10}}$. Thus, for a string scale $M_s\approx{\cal
  O}(1\,{\rm TeV})$, the volume of the internal space $M_6$ needs to
be as large as $V_6M_s^6={\cal O}(10^{32})$.  Second, the
four-dimensional gauge couplings of the strong and weak interactions
are given in terms of the respective volumes $V_p^{P}$, where $P$ runs over the corresponding gauge group factors, as
\begin{equation}
\label{Dpgaugecoupling}
g_{P}^{-2}=(2\pi)^{-1}\ M_s^p\ e^{-\phi_{10}}\ v_p^{(P)}\  .
\end{equation}
Again for a string scale $M_s\approx{\cal O}(1\,{\rm TeV})$ and using
the known values of the strong ($g_3^2/4\pi\approx 0.1$) and the weak
($g_2^2/4\pi\approx g_3^2/12\pi$) gauge coupling constants at the
string scale ($g_2^2/4\pi=\alpha_{\rm EM}/\sin^2\theta_W$, 
$\sin^2\theta_W\approx0.23$, $\alpha_{\rm EM}\approx 1/128$) we can
compute the volumes of the internal cycles, assuming weak string
coupling. To be specific, we choose $g_s=0.2$, and then we obtain
\begin{equation}
M_s^p\ v_p^{(3)}\approx 1\, ,\quad M_s^p\ v_p^{(2)}\approx 3\,.
\label{vol}
\end{equation}
For ${\rm TeV} \ll M_s \alt M_{\rm Pl}$,  $V_6$ and $V_p^{(P)}$'s
are ${\cal O} (1)$ in string units. In general, there are different volumes $V_p^{(P)}$'s for different stacks, and therefore the abelian gauge couplings associated to $U(1)$ symmetries of different D-brane stacks are not equal.

This approach to string model building leads to a variety of low
energy theories including the SM as well as its supersymmetric
extensions. Throughout most of this paper we consider theories which
are non-supersymmetric all the way up to the UV cutoff of the
effective theory; of course the deep UV theory of quantum
gravity may well be supersymmetric. Even though SUSY introduces
special advantages over completely non-SUSY theories, our approach is
distiguished by its simplicity to describe very appealing
phenomenological possibilities that best display the dynamics
involving the extra $U(1)$ symmetries. The study of some aspects of
the supersymmetric version of these models will be postponed until
Sec.~\ref{SEC-V}.

The minimal embedding of the SM particle spectrum requires at least
three brane stacks~\cite{Antoniadis:2000ena} leading to three distinct
models of the type $U(3) \times U(2)\times U(1)$ that were
classified in~\cite{Antoniadis:2000ena, Antoniadis:2004dt}. Only one
of them (model C of~\cite{Antoniadis:2004dt}) has baryon number as
symmetry that guarantees proton stability (in perturbation theory),
and can be used in the framework of TeV-scale strings. Moreover, since
the charge associated to the $U(1)$ of $U(2)$ does not participate
in the hypercharge combination, $U(2)$ can be replaced by  the
symplectic $Sp(1)$ representation of Weinberg-Salam $SU(2)_L$, leading
to a model with one extra $U(1)$ (the baryon number) besides
hypercharge~\cite{Berenstein:2006pk}. 

The SM embedding in four D-brane stacks leads to many more models that
have been classified
in~\cite{Antoniadis:2002qm,Anastasopoulos:2006da}.  In order to make a
phenomenologically interesting choice, herein we focus on models where
$U(2)$ can be reduced to $Sp(1)$. Besides the fact that this reduces
the number of extra $U(1)$'s, one avoids the presence of a problematic
Peccei-Quinn symmetry, associated in general with the $U(1)$ of $U(2)$
under which Higgs doublets are charged~\cite{Antoniadis:2000ena}. To
develop our program in the simplest way, we will work within the
construct of a minimal model, $U(3)_B \times Sp(1)_L \times U(1)_L
\times U(1)_{I_R}$, which has the attractive property of elevating the
two major global symmetries of the SM (baryon number $B$ and lepton
number $L$) to local gauge symmetries~\cite{Cremades:2003qj}. We turn
now to discuss the compelling properties of this model.

\subsection{Standard Model$\bm{^{++}}$}

\begin{figure}[tbp]
\postscript{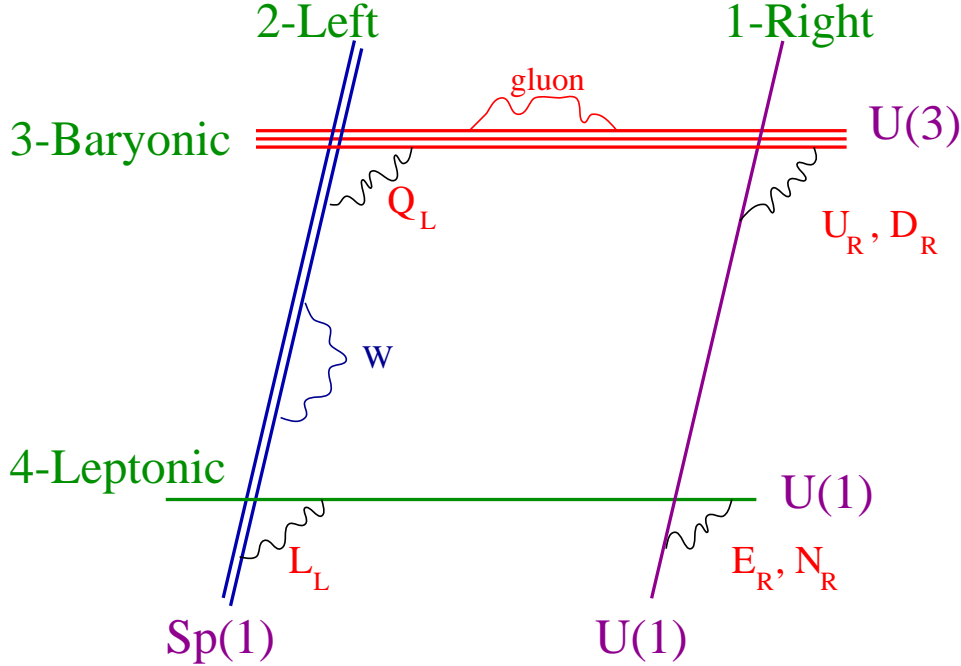}{0.8}
\caption{Pictorial representation of the $U(3)_B \times Sp(1)_L \times U(1)_L \times U(1)_{I_R}$ D-brane model.}
\label{cartoon}
\end{figure}

In this paper we are interested in the minimal 4-stack gauge-extended
sector $U(3)_B \times Sp(1)_L \times U(1)_L \times
U(1)_{I_R}$~\cite{Cremades:2003qj}.  A schematic representation of the
D-brane structure is shown in Fig.~\ref{cartoon} and the brane content
is given in Table~\ref{table-i}. Note that for the $Sp(1)$ stack $P$, the mirror
brane $P^*$ lies on top of $P$. So even though $N_P = 1$, it can be
thought of as a stack of two D6 branes, which give an $Sp(1) \cong
SU(2)$ group under the orientifold projection.  Concretely,
in the bosonic sector the open strings terminating on the stack of
``color'' branes
contain, in addition to the $SU(3)$ octet of gluons 
$$
G^a_{\mu\nu} = \left(\partial_\mu G^a_\nu - \partial_\nu G^a_\mu +
g_3  f^{abc} G_\mu^b  G_\nu^c \right), \quad  i f^{abc}T^a =
\left[T^b,T^c\right], \ T^a \in SU(3) \, ,
$$
an extra $U(1)$ boson $C_\mu$. On the $Sp(1)$ stack the open strings
correspond to the weak gauge bosons
$$
W^a_{\mu\nu} = \left(\partial_\mu W^a_\nu - \partial_\nu W^a_\mu +
 g_2 \epsilon^{abc} W_\mu^b W_\nu^c \right), \quad  i
\epsilon^{abc}\tau^a = \left[\tau^b,\tau^c\right], \ \tau^a \equiv
\sigma^a/2 \in SU(2) \, .
$$
The $U(1)_{I_R}$ D-brane is a terminus for the $B_\mu$ gauge boson,
and there is a third additional $U(1)$ field $X_\mu$ terminating on
the $U(1)_{L}$ brane.  The resulting $U(1)$
content gauges $B$ [with $U(1)_B \subset U(3)_B$], $L$, and a third
additional abelian charge $I_R$ which acts as the third isospin
component of an $SU(2)_R$.  The usual electroweak hypercharge is a
linear combination of these three $U(1)$ charges:
\begin{equation}
Q_Y = c_1 Q_{I_R} + c_3 Q_B + c_4 Q_L   \, ,
\label{hyperchargeY}
\end{equation}
with $c_1 = 1/2$, $c_3 =1/6$, $c_4 = -1/2$, $B=Q_B/3$ and  $L=Q_{L}$.
Alternatively, inverting the above relations, one finds:
\ba
Q_B=3B\quad;\quad Q_{L}=L\quad;\quad Q_{I_R}=2Q_Y-(B-L)  \, .
\label{bb-l}
\ea The chiral particle spectrum from these intersecting branes
consists of six sets (labeled by an index $i = 1, \dots, 6$) of Weyl
fermion-antifermion pairs, whose quantum numbers are given in
Table~\ref{table-ii}. Note that the combination $B-L$ is anomaly free,
while both $B$ and $L$ are anomalous.

\begin{table}
\caption{D-brane content of $U(3)_B \times Sp(1)_L \times U(1)_L \times U(1)_{I_R}$; the
mirror branes $O^*,P^*,Q^*,R^*$ are not shown.}
\begin{center}
\begin{tabular}{cccc}
\hline 
\hline
~~~~Label~~~~ & ~~~~Stack~~~~  & ~~~~Number of Branes~~~~ & ~~~~Gauge Group~~~~ \\
\hline 
$1 \equiv R$ & Right & $N_R =1$ & $U(1)_{I_R}$ \\
$2 \equiv P$ & Left & $N_P =1$ & $Sp(1)_L\cong SU(2)_L$ \\
$3 \equiv Q $ & Baryonic & $N_Q =3$ & ~~~$U(3)_B = SU(3)_C \times U(1)_B$~~~ \\
$4 \equiv O$ & Leptonic & $N_O =1$ & $U(1)_L$ \\
\hline
\hline
\end{tabular}
\end{center}
\label{table-i}
\end{table}

\begin{table}
\caption{Chiral fermion spectrum of the $U(3)_B \times Sp(1)_L \times U(1)_L \times U(1)_{I_R}$ D-brane model.}
\begin{center}
\begin{tabular}{ccccccccc}
\hline
\hline
~~~Label~~~&~~~Fields~~~ & ~~~Sector~~~ &  ~~~$I_{PQ}$~~~ & ~~~Representation~~~ & ~~~$Q_B$~~~ & ~~~$Q_L$~~~ & ~~~$Q_{I_R}$~~~ & ~~~$Q_Y$~~~ \\
\hline
1 & $U_R$ &  $(Q, R^*)$ & 3 &  $( 3,1)$ & $1$ & $\phantom{-}0$ & $\phantom{-} 1$ & $\phantom{-}\frac{2}{3}$  \\[1mm]
2 &  $D_R$ &  $(Q, R)$ & 3 &$( 3,1)$&    $1$ & $\phantom{-}0$ & $- 1$ & $-\frac{1}{3}$   \\[1mm]
3&  $L_L$ & $(O,P)$ & 3 & $(1,2)$&    $0$ &  $\phantom{-}1$ & $\phantom{-}0$ & $-\frac{1}{2}$ \\[1mm]
4&  $E_R$ & $(O,R)$ & 3 & $(1,1)$&   $0$ & $\phantom{-}1$ &  $- 1$ & $- 1$ \\[1mm]
5& $Q_L$ & $(Q,P)$ & 3 & $(3,2)$& $1$ & $\phantom{-}0 $ & $\phantom{-} 0$ & $\phantom{-} \frac{1}{6}$    \\[1mm]
6&  $N_R$&  $(O,R^*)$  & 3 &  $(1,1)$&  $0$ & $\phantom{-}1$ &  $\phantom{-} 1$ & $\phantom{-} 0$ \\
\hline
\hline
\label{table-ii}
\end{tabular}
\end{center}
\end{table}

As mentioned already, the $Q_B$ (gauged baryon number) is
anomalous. This anomaly is canceled by the 4D
version~\cite{Witten:1984dg,Dine:1987xk,Atick:1987gy,Lerche:1987qk,Ibanez:1999it}
of the Green-Schwarz mechanism~\cite{Green:1984sg}. Non anomalous
$U(1)$'s can  acquire masses due to effective six-dimensional
anomalies associated for instance to sectors preserving ${\cal N}=2$
supersymmetry~\cite{Antoniadis:2002cs,Anastasopoulos:2003aj}.\footnote{In
  fact, also the hypercharge gauge boson of $U(1)_Y$ can acquire a
  mass through this mechanism.  In order to keep it massless, certain
  topological constraints on the compact space have to be met.} These
two-dimensional `bulk' masses become therefore larger than the
localized masses associated to four-dimensional anomalies, in the
large volume limit of the two extra dimensions. Specifically for
D$(p+3)$-branes with $p$-longitudinal compact dimensions the masses of
the anomalous and, respectively, the non-anomalous $U(1)$ gauge bosons
have the following generic scale behavior: \ba
{\rm anomalous}~U(1)_a:~~~M_{Z'}&=&g'_aM_s\, ,\label{manom}\\
{\rm non-anomalous}~U(1)_a:~~~M_{Z''}&=&g'_aM_s^3\, V_2\, .  \nonumber\ea Here
$g'_a$ is the gauge coupling constant associated to the group
$U(1)_a$, given by $g'_a\propto g_s/\sqrt{V_p^{(a)}}$ where $g_s$ is
the string coupling and $V_p^{(a)}$ is the internal D-brane
world-volume along the $p$ compact extra dimensions, up to an order
one proportionality constant. Moreover, $V_2$ is the internal
two-dimensional volume associated to the effective six-dimensional
anomalies giving mass to the non-anomalous $U(1)_a$.\footnote{It
  should be noted that in spite of the proportionality of the $U(1)_a$
  masses to the string scale, these are not string excitations but
  zero modes.  The proportionality to the string scale appears because
  the mass is generated from anomalies, via an analog of the
  Green-Schwarz anomaly cancellations: either 4 dimensional anomalies,
  in which case the Green-Schwarz term is equivalent to a
  St\"uckelberg mechanism, or from effective 6 dimensional anomalies,
  in which case the mass term is extended in two more (internal)
  dimensions.}  E.g. for the case of D5-branes, whose common
intersection locus is just 4-dimensional Minkowski-space,
$V_p^{(a)}=V_2$ denotes the volume of the longitudinal,
two-dimensional space along the two internal D5-brane directions.
Since internal volumes are bigger than one in string units to have
effective field theory description, the masses of non-anomalous
$U(1)$-gauge bosons are generically larger than the masses of the
anomalous gauge bosons.\footnote{In \cite{Conlon:2008wa} a different
  (possibly T-dual) scenario with $D7$-branes was investigated. In
  this case the masses of the anomalous and non-anomalous $U(1)$'s
  appear to exhibit a dependence on the entire six-dimensional volume,
  such that the non-anomalous masses become lighter than the anomalous
  ones.}

 The non-anomalous $U(1)_a$ can also remain massless all the
way down to the TeV-scale energy region and grow a mass through a
Higgs mechanism. The absence of a St\"uckelberg mass term for the associated gauge bosons means
that this $U(1)$ gauge symmetry is anomaly free also in six dimensions. In this case a certain topological condition
has to hold, which cannot be read off from the local D-brane quiver, but can only be answered knowing the 6D compact orientifold.
Specifically, just like for the SM gauge symmetry $U(1)_Y$,  the absence of the St\"uckelberg mass term for 
$U(1)_a=c_3^a\,U(1)_B+c_4^a\,U(1)_L+c_1^a\,U(1)_{I_R}$ 
can be 
phrased by the following condition on the homology cycles $\Pi$ and their orientifold images $\Pi'$
of the three $U(1)$ gauge groups:
\begin{equation}
3c_3^a(\Pi_3-\Pi_3')+c_4^a(\Pi_L-\Pi_L')+c_1^a(\Pi_{I_R}-\Pi_{I_R}')=0
\, .
\end{equation}

In what follows we entertain this possibility, having two massless gauge bosons $U(1)_Y$ (associated to the SM hypercharge) and $U(1)_{Y''}$
(associated to a linear combination of anomaly-free
 $I_R$ and $B-L$) and one heavy gauge boson $U(1)_{Y'}$ (associated to an anomalous combination of the three $U(1)$'s).
The classical gauge invariant Lagrangian, obeying the $U(3)_B\times Sp(1)_L\times U(1)_L\times U(1)_{I_R}$
gauge symmetry, can be decomposed as:
\begin{equation}\label{L}
\mathscr{L}_{\rm SM^{^{++}}} = \mathscr{L}_{\rm YM}
+ \sum_{\rm generations} \left(\mathscr{L}_f + \mathscr{L}_{\rm Y} \right) +
\mathscr{L}_{\rm s} +\mathscr{L}_{\rm X}\, ,
\end{equation}
where the terms on the right hand side identify  the
gauge (or Yang-Mills) part, the fermion part, the Yukawa part, the scalar part, and
extra terms from the underlying string theory, respectively.

Electroweak symmetry breaking is achieved through the standard Higgs
doublet $H$. The spontaneous symmetry breaking of the extra non-anomalous $U(1)$
 is attained through an $SU(2)$ singlet scalar field $H''$, which acquires a vacuum
expectation value (VEV) at the TeV scale. The $U(1)$ quantum numbers of
the Higgs sector are given in Table~\ref{table-iii}.

The Yang-Mills
Lagrangian reads:
\begin{equation}
  \mathscr{L}_{\rm YM}
  = -\frac{1}{4} \left(  G^a_{\mu\nu}G_a^{\mu\nu}  +  W^a_{\mu\nu}
    W_a^{\mu\nu} + F^{(1)}_{\mu\nu}F_{(1)}^{\mu\nu} +
    F^{(3)}_{\mu\nu}F_{(3)}^{\mu\nu} + F^{(4)}_{\mu\nu}F_{(4)}^{\mu\nu}
  \right) \,,
  \end{equation}
with the non-Abelian field strengths the same as in the SM,
and the Abelian  $F^{(1)}_{\mu\nu}
  = \partial_\mu B_\nu - \partial_\nu B_\mu,$ $F^{(3)}_{\mu\nu}
  = \partial_\mu C_\nu - \partial_\nu C_\mu$, and $F^{(4)}_{\mu\nu}
  = \partial_\mu X_\nu - \partial_\nu X_\mu$. 

The fermion Lagrangian
  is given by
\begin{eqnarray}
 \mathscr{L}_f  &= &  i \overline{Q_{L}} 
  \gamma_\mu   {\cal D}^\mu  Q_{L} +
  i \overline{U_{R}}  \gamma_\mu  {\cal D}^\mu  U_{R} + i
  \overline{D_{R}}  \gamma_\mu 
  {\cal D}^\mu  D_{R}
  + i \overline{L_{L}}  \gamma_\mu  {\cal D}^\mu \ L_{L}
 +   i \overline{E_{R}}   \gamma_\mu 
  {\cal D}^\mu 
  E_{R} \nonumber \\
& + & i \overline{N_{R}}  \gamma_\mu  {\cal D}^\mu  N_{R} \, ,
\label{elef}
\end{eqnarray}
where
\beq {\cal D}_\mu = \partial_\mu - i g_3 T^a G^a_\mu - i g'_3
Q_B C_\mu - i g_2 \tau^a W^a_\mu - i g'_1 Q_{I_R} B_\mu - i g'_4 Q_{L}
X_\mu 
\label{covderi2}
\eeq are the covariant derivatives with the gauge fields specified
 in the D-brane basis.

\begin{table}
  \caption{Higgs spectrum of the $U(3)_B \times Sp(1)_L \times U(1)_L \times U(1)_{I_R}$ D-brane model.}
\begin{center}
\begin{tabular}{cccccccc}
\hline
\hline
~~~Fields~~~ & ~~~Sector~~~ &  ~~~$I_{PQ}$~~~ & ~~~Representation~~~ & ~~~$Q_B$~~~ & ~~~$Q_L$~~~ & ~~~$Q_{I_R}$~~~ & ~~~$Q_Y$~~~ \\
\hline
$H$ & $(P,R)$ & 1 & $(1,2)$ & $0$ & $\phantom{-}0$ & $\phantom{-}1 $ &
$\frac{1}{2}$  \\ [1mm]
$H''$ & $(O,R)$ & 1 &  $(1,1)$ & $0$ & $-1$ & $-1$  & $0$ \\ 
\hline
\hline
\label{table-iii}
\end{tabular}
\end{center}
\end{table}

The fields $C_\mu,  X_\mu, B_\mu$ are related
to $Y_\mu, Y_\mu{}'$ and $Y_\mu{}''$ by the rotation matrix,
\begin{equation}
\mathbb{R}=
\left(
\begin{array}{ccc}
 C_\theta C_\psi  & -C_\phi S_\psi + S_\phi S_\theta C_\psi  & S_\phi
S_\psi +  C_\phi S_\theta C_\psi  \\
 C_\theta S_\psi  & C_\phi C_\psi +  S_\phi S_\theta S_\psi  & - S_\phi
C_\psi + C_\phi S_\theta S_\psi  \\
 - S_\theta  & S_\phi C_\theta  & C_\phi C_\theta
\end{array}
\right) \,,
\end{equation}
with Euler angles $\theta$, $\psi,$ and
$\phi$~\cite{Anchordoqui:2011ag}. Hence, the covariant derivative
for the $U(1)$ fields in Eq.~(\ref{covderi2}) can be rewritten in terms of
$Y_\mu$, $Y'_\mu$, and $Y''_\mu$ as follows
\begin{eqnarray}
\CD_\mu & = & \partial_\mu -i Y_\mu \left(-S_\xt g'_1 Q_{I_R} + C_\theta S_\psi  g'_4  Q_{L} +  C_\theta C_\psi g'_3 Q_B \right) \nonumber \\
 & - & i Y'_\mu \left[ C_\theta S_\phi  g'_1 Q_{I_R} +\left( C_\phi C_\psi + S_\theta S_\phi S_\psi \right)  g'_4 Q_{L} +  (C_\psi S_\theta S_\phi - C_\phi S_\psi) g'_3 Q_B \right] \label{linda} \\
& - & i Y''_\mu \left[ C_\theta C_\phi g'_1 Q_{I_R} +  \left(-C_\psi S_\phi + C_\phi S_\theta S_\psi \right)  g'_4  Q_{L} + \left( C_\phi C_\psi S_\theta + S_\phi S_\psi\right) g'_3 Q_B \right]   \, .  \nonumber
\end{eqnarray}
Now, by demanding that $Y_\mu$ has the
hypercharge $Q_Y$ given in Eq.~\er{hyperchargeY}  we  fix the first column of the rotation matrix $\mathbb{R}$
\begin{equation}
\bay{c} C_\mu \\  X_\mu \\ B_\mu
\eay = \left(
\begin{array}{lr}
  Y_\mu \,  c_3 g_Y/g'_3& \dots \\
  Y_\mu \,  c_4 g_Y/g'_4 & \dots\\
   Y_\mu \, c_1 g_Y/g'_1 & \dots
\end{array}
\right) \, ,
\end{equation}
and we determine the value of the two associated Euler angles
\begin{equation}
\theta = {\rm -arcsin} [c_1 g_Y/g'_1]
\label{theta}
\end{equation}
and
\begin{equation}
\psi = {\rm arcsin}  [c_4 g_Y/ (g'_4 \, C_\theta)] \, .
\label{psi}
\end{equation}
The couplings $g'_1$ and $g'_4$ are related through the orthogonality
condition, $P(g_Y,g'_1,g'_3,g'_{4}) =0$, yielding
\begin{equation}
 \left(\frac{c_4}{ g' _4} \right)^2  = \frac{1}{g_Y^2} - \left(\frac{c_3}{g'_3} \right)^2  - \left(\frac{c_1}{g'_1}\right)^2  \, ,
\label{orthogonality25}
\end{equation}
with $g'_3$ fixed by the relation $g_3 (M_s) = \sqrt{6} \, g'_3
(M_s)$~\cite{Anchordoqui:2011eg}.  Next, by demanding that $Y''$
couples to a linear combination of anomaly-free $I_R$ and $B-L$ we
determine the third Euler angle
\begin{equation}
\tan \phi = - S_\theta \frac{3 \ g'_3 \ C_\psi + g'_4 \ S_\psi}{3 \ g'_3 \ 
  S_\psi - g'_4 \ C_\psi}  \  .
\label{tanfi}
\end{equation}

In the $(Y,Y',Y'')$ basis, $Y$ and $Y''$ are coupled to anomaly-free
currents while the anomaly of the current associated to $Y'$ is
cancelled by the generalized Green-Schwarz mechanism. As a result,
$Y'$ acquires a mass of order of the string mass $M_s$, c.f.\
Eq.~(\ref{manom}).  Higgs VEVs will generate additional mass terms for
$Y'$, introducing also some small mixing with other gauge gauge
bosons, of order $({\rm TeV}/M_s)^2$. From now on, we neglect such
small effects and take $Y' \simeq Z'$.

 The Yukawa interactions are given by
\begin{equation}\label{Yukawas}
\mathscr{L}_{\rm Y} =  - Y_d \left(\overline{Q_L} H \right)D_R -Y_u \left(
  \overline{Q_L} i \sigma^2 H^* \right) U_R - Y_e \left(\overline{L_L}
  H\right)E_R  - Y_{N} \left(\overline{L_L} i \sigma^2 H^* \right) N_R  +
{\rm h.c.},
\end{equation}
where the Yukawa couplings $Y_i$ are matrices in flavor space.
Note that unlike in the supersymmetric case, a single Higgs vacuum expectation value will generate masses
for up and down quarks.\footnote{$i\sigma_2H^*$ transforms in the fundamental representation of $SU(2)$.}

Note that with the charge assignments of Tables~\ref{table-ii} and
\ref{table-iii} there are no dimension 4 operators involving $H''$
that contribute to the Yukawa Lagrangian. This is very important since
$H''$ carries the quantum numbers of right-handed neutrino and its VEV
breaks lepton number. However, this breaking can affect only
higher-dimensional operators which are suppressed by the high string
scale, and thus there is no phenomenological problem with experimental
constraints for $M_s$ higher than $\sim 10^{14}$ GeV.

The scalar Lagrangian is
\begin{equation}\label{new-scalar_L}
\mathscr{L}_s=\left( {\cal D}^{\mu} H\right) ^{\dagger} {\cal D}_{\mu}H + 
\left( {\cal D}^{\mu} H'' \right) ^{\dagger} {\cal D}_{\mu}H'' - V(H,H'' ) \, ,
\end{equation}
with the potential
\beq
V\left(H, H'' \right) = \mu^2 \left| H \right|^2 +{ \mu'}^2 \left| H''
\right|^2 + \lambda_1 \left| H \right|^4 + \lambda_2 \left| H''
\right|^4 + \lambda_3 \left| H \right|^2 \left| H'' \right|^2 \ .
\label{higgsV}
\eeq 
The Higgs VEVs obtained after minimizing this potential will be denoted as
\beq
\langle\,  H\,\rangle =
 \begin{pmatrix}
  0 \\
  v  \\
 \end{pmatrix}  
\quad {\rm and} \quad
\langle H''\rangle  =  v''  \, .
\label{s16} 
\eeq  
The kinetic terms of the Higgs fields in (\ref{new-scalar_L})
give masses to the various gauge bosons.

At this point, we identify the photon $A_\mu$ and weak
force mediators $W^+_{\mu},W^-_{\mu}, \overline Z_{\mu}$ performing the usual
Weinberg rotation \beq
\begin{pmatrix}
  A_\mu \\
  \overline Z_\mu \\
  W^+_\mu \\
  W^-_\mu \\
 \end{pmatrix} = \begin{pmatrix}
  \phantom{-}C_{\theta_W} & S_{\theta_W} & 0 & \phantom{-}0  \\
  -S_{\theta_W} & C_{\theta_W} & 0 & \phantom{-}0 \\
  \phantom{-}0 & 0 & 1/\sqrt{2} & \phantom{-}i/\sqrt{2} \\
 \phantom{-} 0 & 0 & 1/\sqrt{2} & -i/\sqrt{2} \\
 \end{pmatrix} 
 \begin{pmatrix}
  Y_\mu \\
  W^3_\mu \\
  W^1_\mu \\
  W^2_\mu \\ 
 \end{pmatrix} \, ;
\eeq
this gives
\begin{eqnarray}
 {\cal D}_\mu & = & \partial_\mu  -  \frac{i}{2} \, g_2 \, \sigma^- W^+_\mu - \frac{i}{2} \, g_2 \, \sigma^+ W^-_\mu  - i g_2 \, \cos \theta_W \, \left(\sigma^3/2 - Q_Y \tan^2 \theta_W \right) \overline Z_\mu  -   i g_2 \sin \theta_W  \nonumber \\
& \times &\left(\sigma^3/2 + Q_Y \right) A_\mu  -  i g_{Y'} Q_{Y'} Z'_\mu - i g_{Y''} Q_{Y''} Y''_\mu \,,
\label{35}
\end{eqnarray}
with $\sigma^{\pm} = \left(\sigma^1 \pm i \sigma^2\right)/2$ ,
$g_Y/g_2 = \tan\theta_W$. From (\ref{linda}) and (\ref{35}) we define
\begin{eqnarray}
Q_Y H & = & H/2 \,, \nonumber \\ 
g_{Y'} Q_{Y'} H & = & (g'_1 C_\theta S_\phi ) H \,, \nonumber \\ 
g_{Y''} Q_{Y''} H
 & = & (g'_1 C_\theta C_\phi) H \, , \nonumber \\
 Q_Y H'' & = & 0 \,, \nonumber \\
g_{Y'} Q_{Y'} H'' & = &
-[g'_1 C_\theta S_\phi + g'_4(C_\phi C_\psi + S_\theta S_\phi
S_\psi)]H'' \,, \nonumber \\
g_{Y''} Q_{Y''} H'' & = & -(g'_1C_\theta C_\phi +
g'_4 [C_\phi S_\theta S_\psi - C_\psi S_\phi)]H'' \, .
\end{eqnarray}
The Higgs kinetic terms of Eq.(\ref{new-scalar_L})
 together with the
Green-Schwarz mass term, $\frac{1}{2} {M'}^2 Z'_\mu Z'^\mu$, lead to
\begin{equation}
\mathscr{B} = [{\cal D}^\dagger_\mu \left( 0 \ v\right)] \left[{\cal
    D}^\mu \left(\begin{matrix} 0 \\ v \end{matrix}\right) \right] +
({\cal D}_\mu v'')^\dagger({\cal D}^\mu v'') + \frac{1}{2} {M'}^2
Z'_\mu Z'^\mu  \, .
\end{equation}
Expanded this gives
\begin{eqnarray}
  \mathscr{B} & = & \frac{1}{4} (g_2\, v)^2 W^+_\mu W^{-\mu} + 
 \frac{1}{4} (g_2 v)^2  C^{-2}_{\theta_W} \, \overline Z_\mu \overline Z^{\mu} 
  +  g'_1  C_\theta \left(S_\phi Z'_\mu + C_\phi Y''_\mu\right) g_2 \ v^2 C_{\theta_W}^{-1} \overline Z^{\mu} \nonumber \\
  & + & {v''}^2\left\{ g'_1 C_\theta (S_\phi \, Z'_\mu + C_\phi \, Y''_\mu) + g'_4 \left[ (C_\phi C_\psi + S_\theta S_\phi S_\psi) \, Z'_\mu + S_\psi S_\theta C_\phi \, Y''_\mu \right] \right\}^2  \nonumber \\
  & + & 
  (g'_1 v \ C_\theta)^2 \left(S_\phi Z'_\mu + C_\phi
    Y''_\mu\right)\left(S_\phi Z'^\mu + C_\phi Y''^\mu\right)
  + \frac{1}{2} {M'}^2 Z'_\mu Z'^\mu \nonumber \\
  & \simeq & \frac{1}{4} (g_2\, v)^2 W^+_\mu W^{-\mu} + 
 \frac{1}{4} (g_2 v)^2  C^{-2}_{\theta_W} \, \overline Z_\mu \overline Z^{\mu} 
 +  g'_1  C_\theta  C_\phi Y''_\mu  \, g_2 \ v^2 C_{\theta_W}^{-1} \overline Z^{\mu} \nonumber \\
  & + & {v''}^2\left( g'_1 C_\theta  C_\phi \, Y''_\mu + g'_4  S_\psi S_\theta 
C_\phi \, Y''_\mu  \right)^2  + 
  (g'_1 v \ C_\theta C_\phi)^2 
    Y''_\mu Y''^\mu
  + \dots 
\label{DB1}
\end{eqnarray}
where the  omitted terms pertain only to the $Z'$ couplings at the
string scale. Recall that we have taken $M' \sim M_s$ and therefore $Z'$
decouples from the low energy physics. By inspection of (\ref{DB1}) we
immediately recognize the $W^\pm$ masses and the usual tree level
formula for the mass of the $Z$ particle in the electroweak theory,
$\overline M_Z^2 = (g_2^2 v^2 + g_Y^2 v^2)/2$, before mixing.

Now, we use the relation $g'_1 S_\theta = g'_4 C_\theta S_\psi$ to
conveniently rewrite (\ref{DB1}) as
\begin{eqnarray}
\mathscr{B} & \simeq &\frac{1}{4} \, (g_2 v)^2 \, W^+_\mu W^{-\mu} + \frac{1}{4} (g_2 v)^2  C^{-2}_{\theta_W} \,  \overline Z_\mu \overline Z^{\mu} +   
g'_1 v C_\theta \, C_\phi \, g_2 v C^{-1}_{\theta_W} \,  Y''_\mu \overline Z^{\mu} +  \left(\frac{v'' g'_1
    C_\phi}{C_\theta}\right)^2 \nonumber \\ 
& \times & \left( 1+ \left(\frac{v}{v''}
    C_\theta^2\right)^2\right) Y''_\mu Y''^\mu +... \nonumber \\
& \simeq & \frac{1}{4} \, (g_2 v)^2 \, W^+_\mu W^{-\mu} +
\left(\frac{v'' g'_1 C_\phi}{C_\theta}\right)^2 \left[ 1+ \left(\frac{v}{v''} C_\theta^2\right)^2\right] \left\{ Y_\mu'' + \frac{g'_1 \ C_\theta^3 \ C_\phi \ g_2 \ v^2 \ C_{\theta_W}^{-1} \ \bar{Z}_\mu }{2 \left(v'' g'_1 C_\phi\right)^2 \left[ 1+ \left(\frac{v}{v''} C_\theta^2\right)^2\right]}  \right\}^2 \nonumber \\
& + & \left\{ \frac{1}{4} (g_2 v)^2  C^{-2}_{\theta_W} \,
- \frac{g'_1 \ C_\theta^3 \ C_\phi \ g_2 \ v^2 \ C_{\theta_W}^{-1} \ }{2 \left(v'' g'_1 C_\phi\right)^2 \left[ 1+ \left(\frac{v}{v''} C_\theta^2\right)^2\right]} \right\}\  \bar{Z}_\mu \ \bar{Z}^\mu + ... \ .
\label{BH2}
\end{eqnarray}
Finally, if we make the expansion around $v/v'' \ll
1$, the $\overline Z_\mu {Y''}^\mu$ mass matrix is render diagonal and we obtain
the desired expression for the mass terms \beq \mathscr{B}
=\left(\frac{g_2 v}{2}\right)^2  \, W^+_\mu W^{-\mu} + \left(\frac{g_2
    v}{2  C_{\theta_W}}\right)^2  Z_\mu Z^{\mu} + \left(\frac{g'_1\, C_\phi \, v''}{ C_\theta}\right)^2 \, Z''_\mu Z''^\mu  
+ {\cal O} \left(\left(\frac{v}{v''}\right)^2\right)\, , \eeq where $Z'' \simeq Y'' + {\rm small \
  corrections}$.

In principle, in addition to the orthogonal field mixing induced by
identifying anomalous and non-anomalous $U(1)$ sectors, there may be
kinetic mixing between these sectors. However, in models where there
is only one $U(1)$ per stack of D-branes, the relevant kinetic mixing
is between $U(1)$'s on different stacks, and hence involves loops with
fermions at brane intersection. Such loop terms are typically down by
${g'_i}^2/16 \pi^2 \sim 0.01$~\cite{Dienes:1996zr}.\footnote{The major
  effect of the kinetic mixing is in communicating SUSY breaking from
  a hidden $U(1)$ sector to the visible sector, generally in
  modification of soft scalar masses. For a comprehensive review of
  experimental limits on the mixing, see~\cite{Abel:2008ai}.}  By
inspection of Table~\ref{table-ii}  the charges
$Q_B$, $Q_L$, and $Q_{I_R}$ are mutually orthogonal in the fermion
space, {\em i.e} $\sum_fQ_{i,f}Q_{j,f}=0$ for $i\neq j$. This will
maintain the othogonality relation $P=0$ to one loop without inducing
kinetic mixing~\cite{Anchordoqui:2011eg}. The charges
  assigned to $H''$ (see Table~\ref{table-iii}) will violate the orthogonality
  condition. However, the $H''$ only contributes at the 0.9\% level to
  the running of $g'_1$ from the string scale to the TeV scale, and
  about 0.3\% to the running of $g'_4$. These are of the same order as
  the two loop contributions from the fermion sector, so we may
  consistently ignore the nonorthogonality introduced by $H''$ in the
  context of one loop considerations.

\section{LHC Phenomenology}
\label{SEC-III}

In this section we discuss the discovery potential of the $Z''$
resonance at the LHC.  Before proceeding, we summarize the lessons
learned thus far.  The initially free parameters of the model consist
of three couplings $g'_1,\, g'_3,\, g'_{4}.$ These are augmented by
three Euler angles to allow for a field rotation to coupling diagonal
in hypercharge. This diagonalization fixes two of the angles and the
orthogonal nature of the rotation introduces one constraint on the
couplings $P(g_Y,g'_1,g'_3,g'_{4}) =0$. The baryon number coupling
$g'_3$ is fixed to be $\sqrt{1/6}$ of the non-abelian $SU(3)$ coupling
at the scale of $U(N)$ unification, and is therefore determined at all
energies through RG running.  In what follows, we take $M_s =
10^{14}~{\rm GeV}$ as a reference point for running down the $g'_3$
coupling to the TeV region that is {\em ignoring mass threshold
  effects of stringy states}. This yields $g'_3 (M_s)= 0.231$. We have
checked that the running of the $g'_3$ coupling does not change
significantly within the LHC range, for different values of the string
scale. This leaves one free angle and two couplings with one
constraint. Equation~(\ref{tanfi}) fixes the third Euler angle.  To
comply with these assignments and ensure perturbativity of $g'_4$
between the TeV scale and the string scale we find from
(\ref{orthogonality25}) that $g'_1 > 0.4845$. We also take $g'_1 \alt
1$ in order to ensure perturbativity at the string scale.

We first consider the case with $g'_1 (M_s) \simeq 1$. This leads to
$\psi (M_s) = -1.245$, $\theta (M_s) = -0.217$, $\phi (M_s) =
-0.0006$, and $g'_4 (M_s) = 0.232$. Substituting our fiducial values
in (\ref{linda}) we find the non-anomalous $U(1)$ vector bosons couple
to currents
\begin{eqnarray}
J_Y & = & 2.1 \times 10^{-1}~Q_{I_R} + 2.1 \times 10^{-1}~(B-L) \nonumber \\     J_{Y''} & = & 9.8 \times 10^{-1}~Q_{I_R}  - 4.7 \times 10^{-2}~(B-L)  \, ,
\end{eqnarray}
at the string scale. Next, we run the couplings down to the TeV
region. A very important point is that the couplings that are running
are those of the $U(1)$ fields; hence the $\beta$ functions receive
contributions from fermions and scalars, but not from gauge bosons.
The one loop correction to the various couplings are
\begin{equation} \frac{1}{\alpha_Y(Q)} = \frac{1}{\alpha_Y (M_s)} -
  \frac{b_Y}{2\pi} \, \ln(Q/M_s) \,, \end{equation} \begin{equation}
  \frac{1}{\alpha_i(Q)} = \frac{1}{\alpha_i (M_s)} - \frac{b_i}{2\pi}
  \, \ln(Q/M_s) \,, \label{RGbi} \end{equation} where \begin{equation}
  b_i = \frac{2}{3} \, {\sum_f} \, Q_{i,f}^2 \, + \frac{1}{3} \,
  {\sum_s} \, Q_{i,s}^2, \end{equation} with $f$ and $s$ indicating
contribution from fermion and scalar loops, respectively. Setting $Q =
4~{\rm TeV}$, from (\ref{RGbi}) we obtain: $g'_1 = 0.406$, $g'_3 =
0.196$, $g'_4 = 0.218$, $\theta = -0.466$, $\psi = -1.215$, and $\phi
= -0.0003$.  This leads to 
\begin{eqnarray}
  J_{Y}& = & 1.8 \times 10^{-1}~Q_{I_R} + 1.8 \times 10^{-1}~(B - L) \nonumber \\
  J_{Z''} & = & 3.6 \times 10^{-1}~Q_{I_R}  - 9.2 \times 10^{-2}~(B-L)  \, ,
\end{eqnarray}
where we have assumed that $H''$ has developed its VEV.  Since ${\rm
  Tr}~[Q_{I_R} \, B] = {\rm Tr}~[Q_{I_R} L] = 0$, the $Z''$ decay
width is given by
\begin{eqnarray}
\Gamma_{Z''} & = & \Gamma_{Z'' \to Q_{I_R}}  + \Gamma_{Z'' \to B-L} \nonumber \\
& \propto & ( 1.4 \times 10^{-1})^2 \, {\rm Tr}[Q_{I_R}^2]  + (9.2\times 10^{-2})^2 {\rm Tr}\left[(B-L)^2 \right] \nonumber \\
& = & 1.0 \times 10^{0} + 4.5 \times 10^{-2} \, .
\end{eqnarray}
Thus, the corresponding branching fractions are BR~$Z'' \to Q_{I_R} =
0.959$ and BR~$Z'' \to B-L = 0.041$. Though not relevant for LHC
phenomenology, a straightforward calculation shows that $Z'$ is very
nearly diagonal in $B$, with BR $Z' \to B = 0.946$ and BR $Z' \to L =
 0.054$. Of course, since the quiver construction has each particle
straddling two adjacent branes, there can be considerable variation in
decay channels particle by particle. This is evident in
Table~\ref{table-iv}.\footnote{The physical couplings of the $Z''$ to
  fermions fields given in Table~\ref{table-iv} are consistent with
  the bounds presented in~\cite{Williams:2011qb} from a variety of
  experimental constraints.}  The dominance of $B$ for the $Z'$ decay
channel and $I_R$ for the $Z''$ decay channel is valid after averaging
over decay channels.\footnote{An analogue is in the SM. The $Z$
  couples to a current $J_Z \propto T_3 - \tan^2 \theta_W
  \frac{Y}{2}$, where $Q = T_3 - \frac{Y}{2}$. In this case, $\sum
  (\frac{Y}{2})^2 = \frac{17}{6}$ and ${\rm Tr} [T_3^2] = 2$; we have
  ${\rm BR}~Z \to T_3 : {\rm BR}~Z \to \frac{Y}{2} = 2 :\frac{17}{6}
  \tan^4 \theta_W = 2 : 0.25 = 8:1$. However, this certainly does not
  hold particle by particle; e.g., for the neutrino electron doublet:
  $\Gamma_{Z \to \nu} \propto (1 + \tan^2 \theta_W)^2 \sim 1.7$,
  whereas $\Gamma_{Z\to e} \propto (1 - \tan^2 \theta_W)^2 \sim 0.5$.}

\begin{figure}[tbp] 
\begin{minipage}[t]{0.49\textwidth}
    \postscript{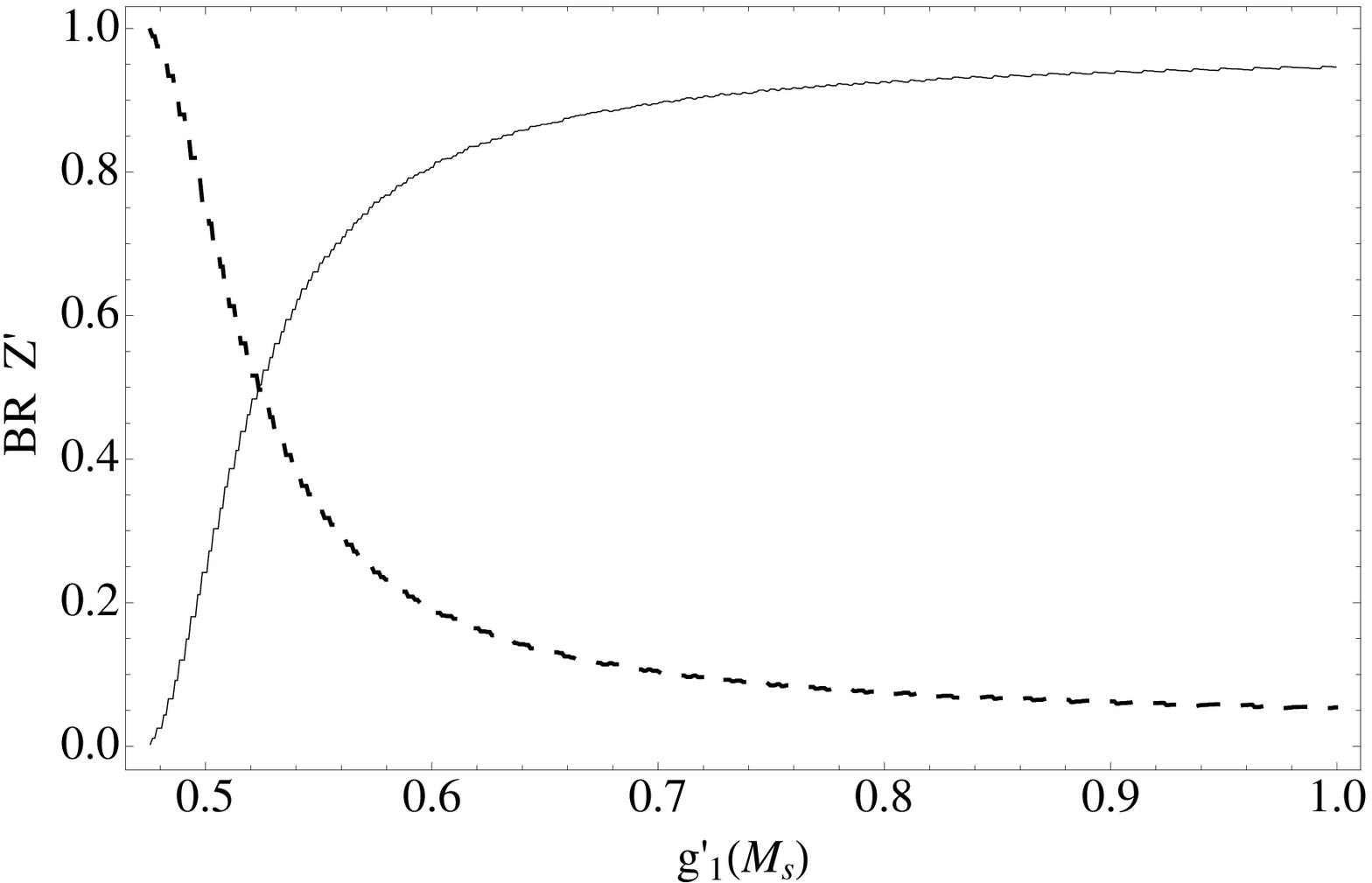}{0.99} \end{minipage}
  \hfill \begin{minipage}[t]{0.49\textwidth}
    \postscript{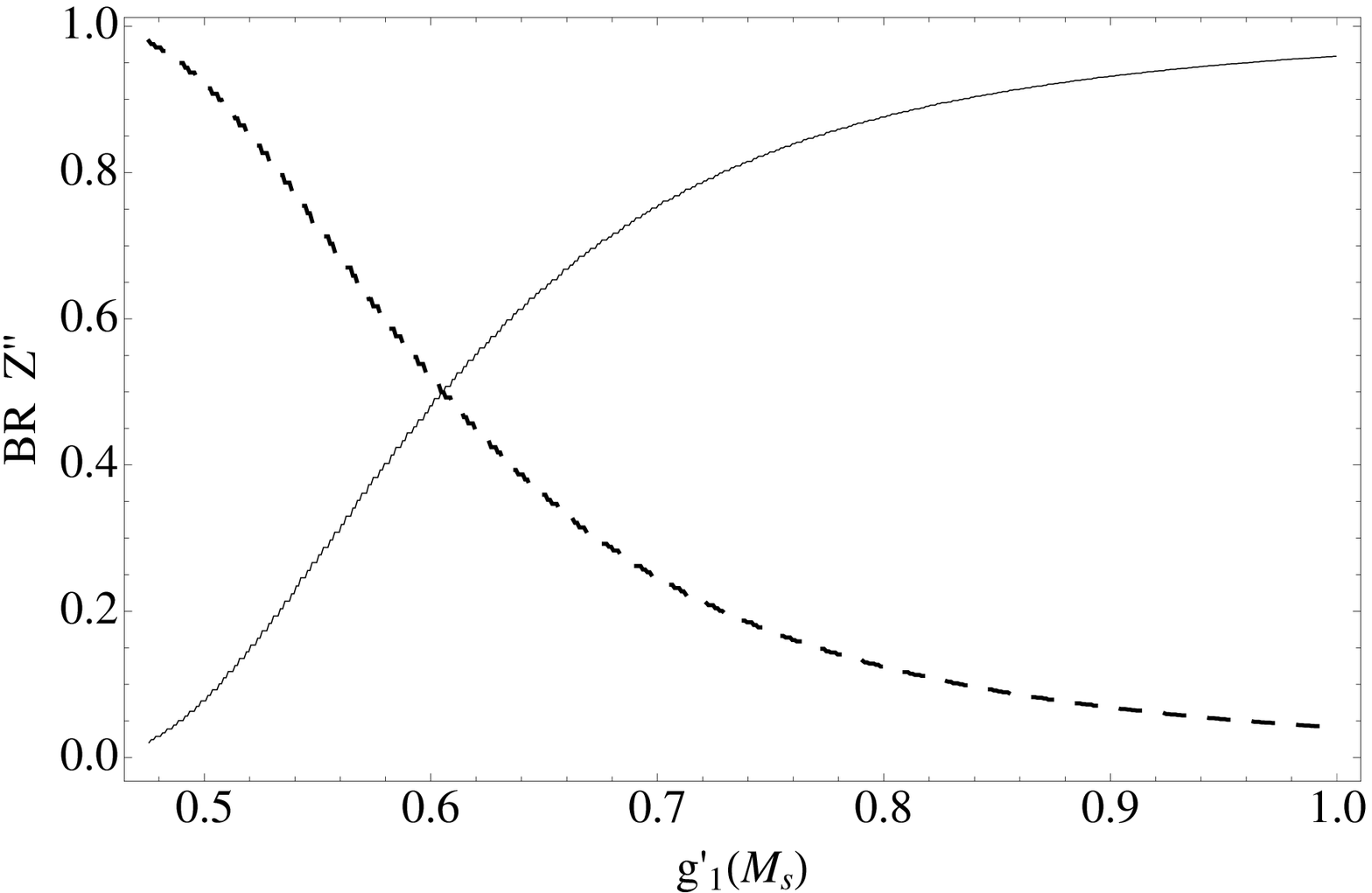}{0.99} \end{minipage} 
  \caption{Branching fractions of $Z'$ (left) and $Z''$ (right) as a function of
    $g'_1(M_s)$. The solid lines denote the branching into $B$ (left) and $I_R$ (right). The dashed lines denote the
    branching into $L$ (left) and B-L (right).}
\label{sm++_BR} 
\end{figure}

Now, duplicating the procedure for $g'_1(M_s) = 0.4845$ we obtain
\begin{equation}
\begin{tabular}{ccccccccccc} 
 BR& $Z' \to B$ & : & BR &$Z' \to L$ & : & BR &$Z'' \to Q_{1R}$  &:&
 BR &$Z'' \to B-L$   \\
  & $0.066$ & : & & $0.934$ & : & & $0.039$ &  : & & $0.961$ \, .
\end{tabular}
\end{equation}
The chiral couplings of $Z'$ and $Z''$ gauge bosons which are mostly
$L$ and $B-L$, respectively are given in Table~\ref{newtable}. The
variation of the branching fractions within the allowed range of
$g'_1(M_s)$ is shown in Fig.~\ref{sm++_BR}.

The LHC discovery potential for $Z''$ gauge boson as a mass peak above
a small background in the reactions $pp \to Z'' \to jj $ and $pp \to
Z'' \to \ell^+\ell^-$, with $\ell = e, \mu,$ is well known. The required
luminosity to discover a $Z''$ basically depends only on its cross
section, and therefore on its mass and couplings. Experimental effects
due to mass resolution are known to result in an only minor reduction
of the sensitivity.

\begin{figure}[tbp]
\begin{minipage}[t]{0.49\textwidth}
    \postscript{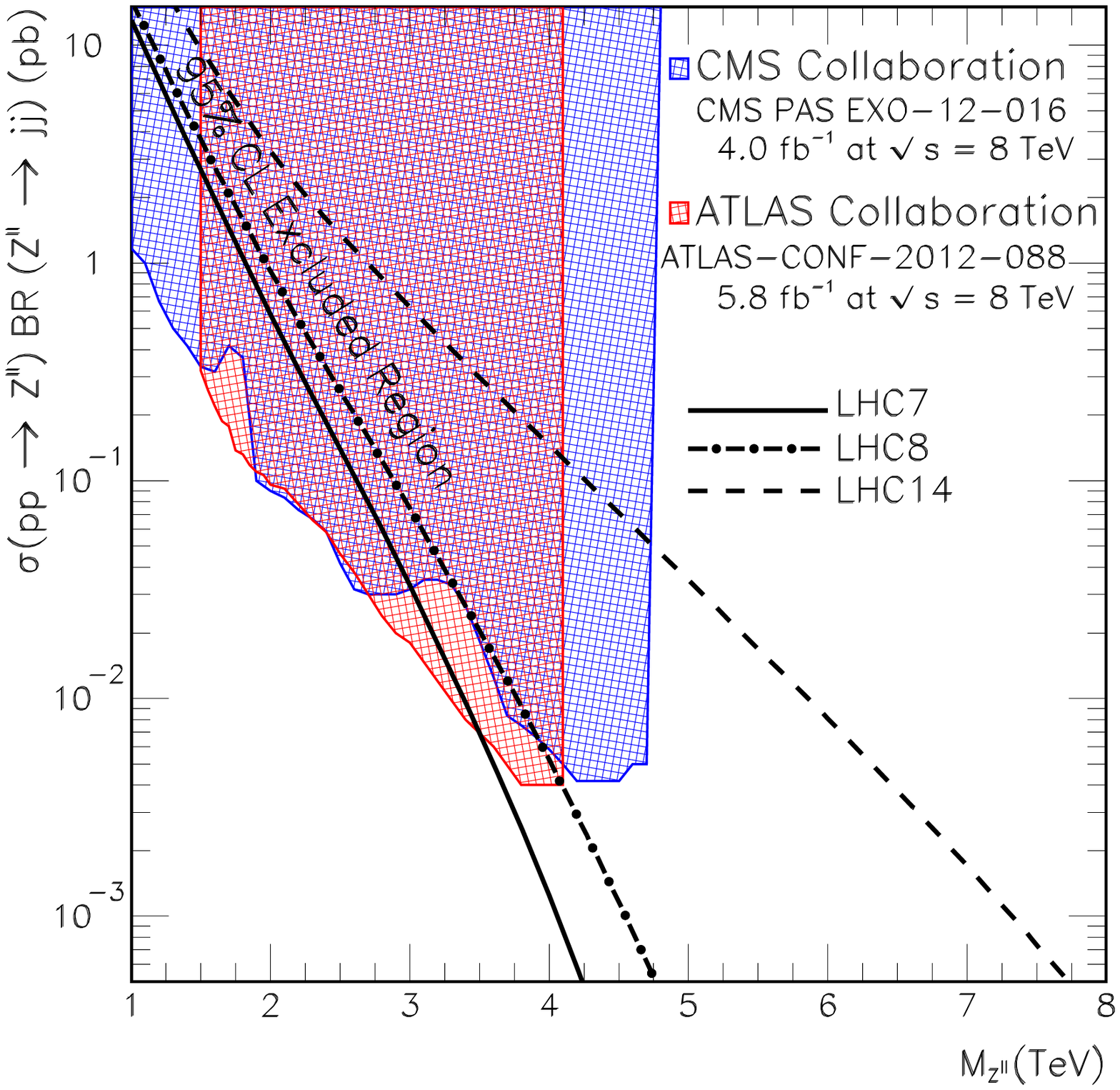}{0.99} \end{minipage}
  \hfill \begin{minipage}[t]{0.49\textwidth}
    \postscript{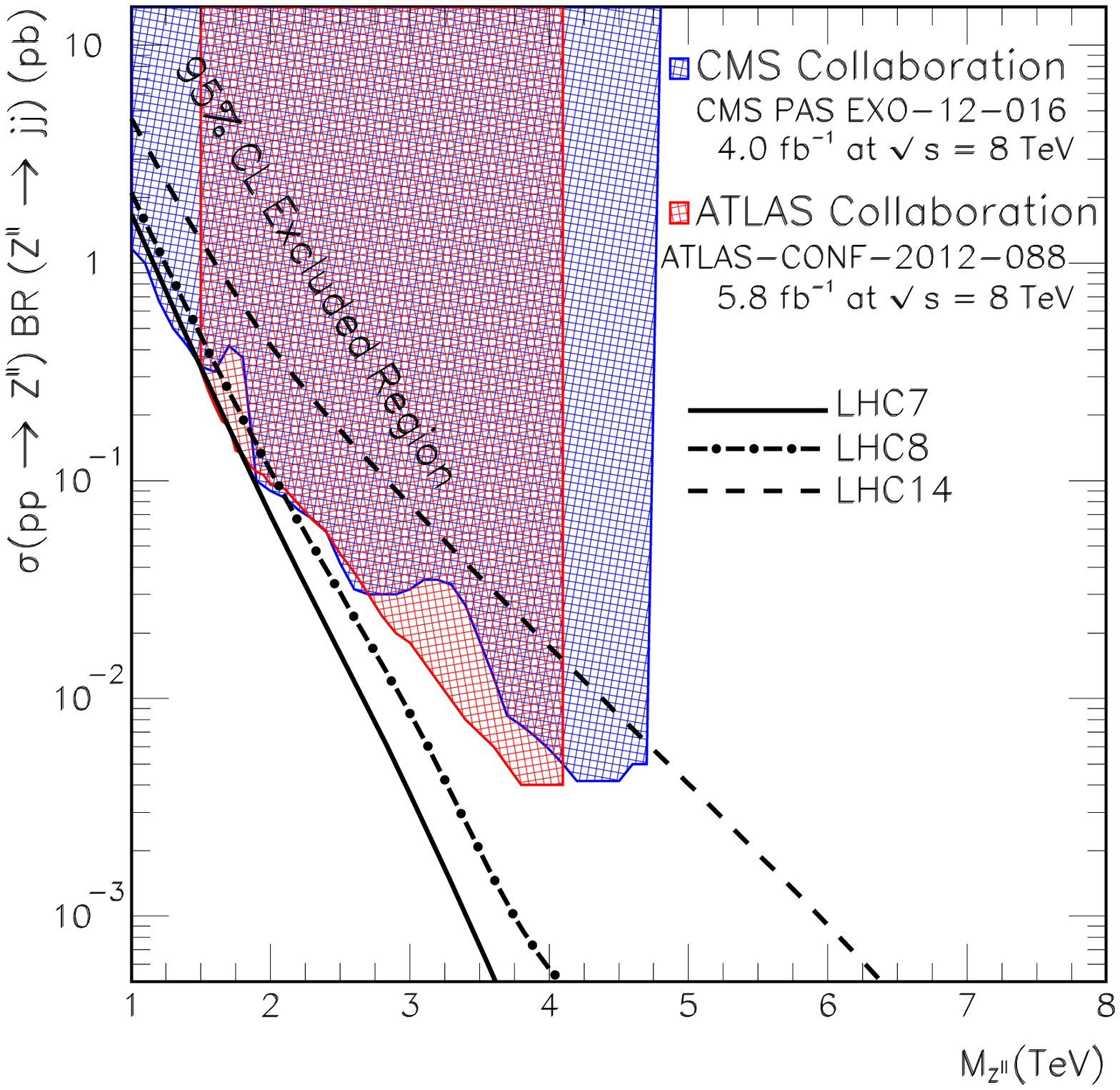}{0.99} \end{minipage} 
  \caption{Comparison of the (pre-cut) total cross section for the
    production of $p p \to Z'' \to jj$ with the 95\% CL upper limits
    on the production of a gauge boson decaying into two jets as
    reported by the CMS and ATLAS collaborations (corrected by
    acceptance). For isotropic decays (independently of the
    resonance), the acceptance for the CMS detector has been reporetd
    to be ${\cal A} \approx 0.6$. The ATLAS acceptance ranges from
    11\% to 54\% varying from 1~TeV to 4.25~TeV, and is never lower
    than 48\% for masses above 2~TeV. The case in which $Z''$ is
    mostly diagonal in $I_R$ is shown in the left panel and the case
    in which it is mostly $B-L$ in the right panel.}
\label{fig:CMS} 
\end{figure}

Using a data set of $pp$ collisions at $\sqrt{s} = 8~{\rm TeV}$, with
an integrated luminosity of $4.0~{\rm fb}^{-1}$, the CMS Collaboration
has searched for narrow resonances in the dijet invariant mass
spectrum~\cite{cms_dijet8tev}. Each event in the search is
required to have its two highest-$p_T$ jets with (pseudorapidity)
$|\eta_j| < 2.5$. The acceptance ${\cal A}$ of selection requirements
is reported to be $\approx 0.6$.  The spectra are consistent with SM
expectations and thus upper limits on the cross section times
branching fraction for $Z''$ into two jets have been set. Similar
upper limits have been obtained by the ATLAS Collaboration using
$5.8~{\rm fb}^{-1}$ of data collected at $\sqrt{s} = 8~{\rm
  TeV}$~\cite{atlas_dijet8tev}. These results, which are display in
Fig.~\ref{fig:CMS}, extend previous exclusion limits from
LHC7~\cite{Khachatryan:2010jd,Chatrchyan:2011ns,Aad:2011fq,cms_dijet7tev,atlas_dijet7tev}.

\begin{figure}[tbp]
\begin{minipage}[t]{0.49\textwidth}
    \postscript{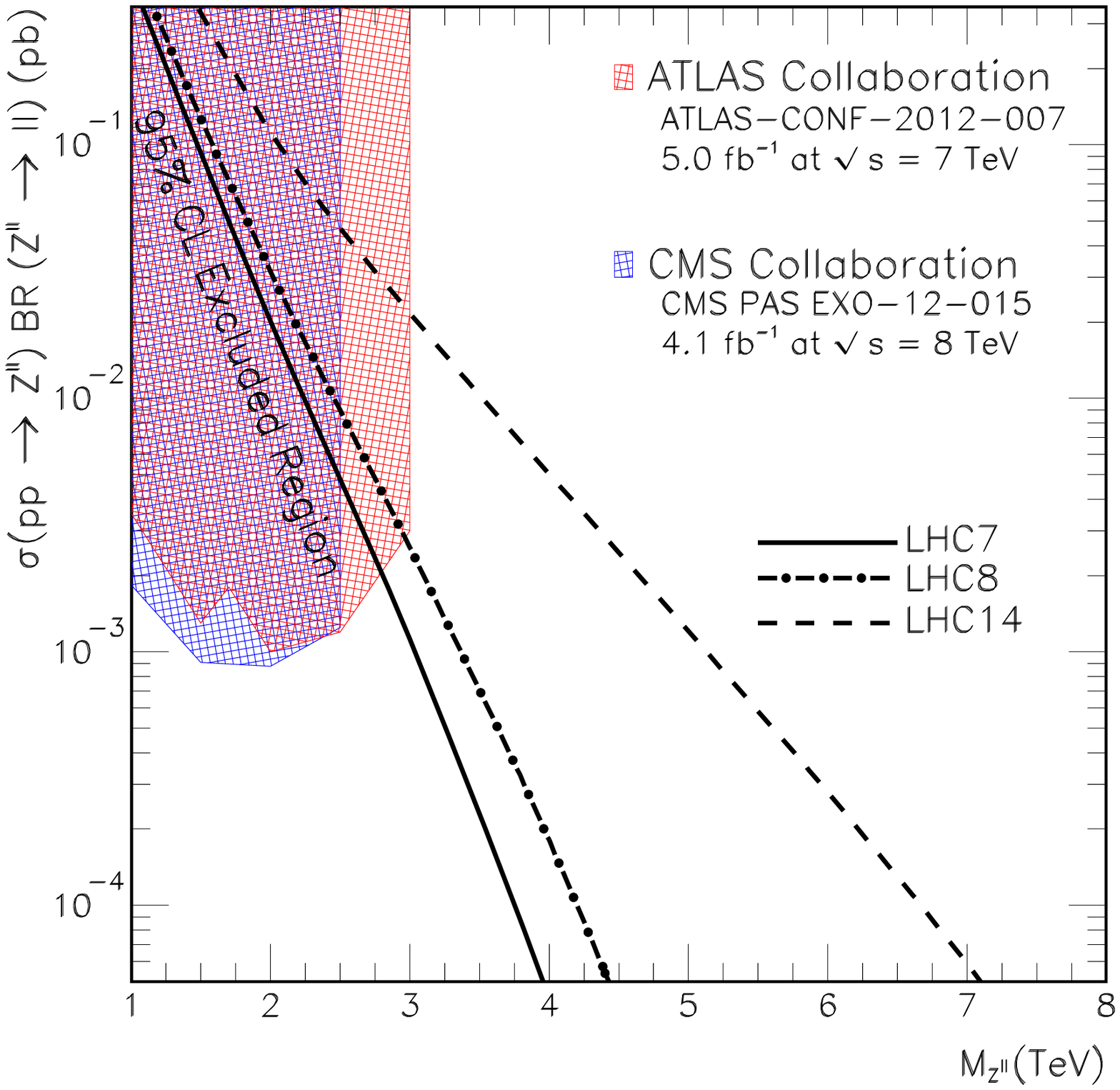}{0.99} \end{minipage}
  \hfill \begin{minipage}[t]{0.49\textwidth}
    \postscript{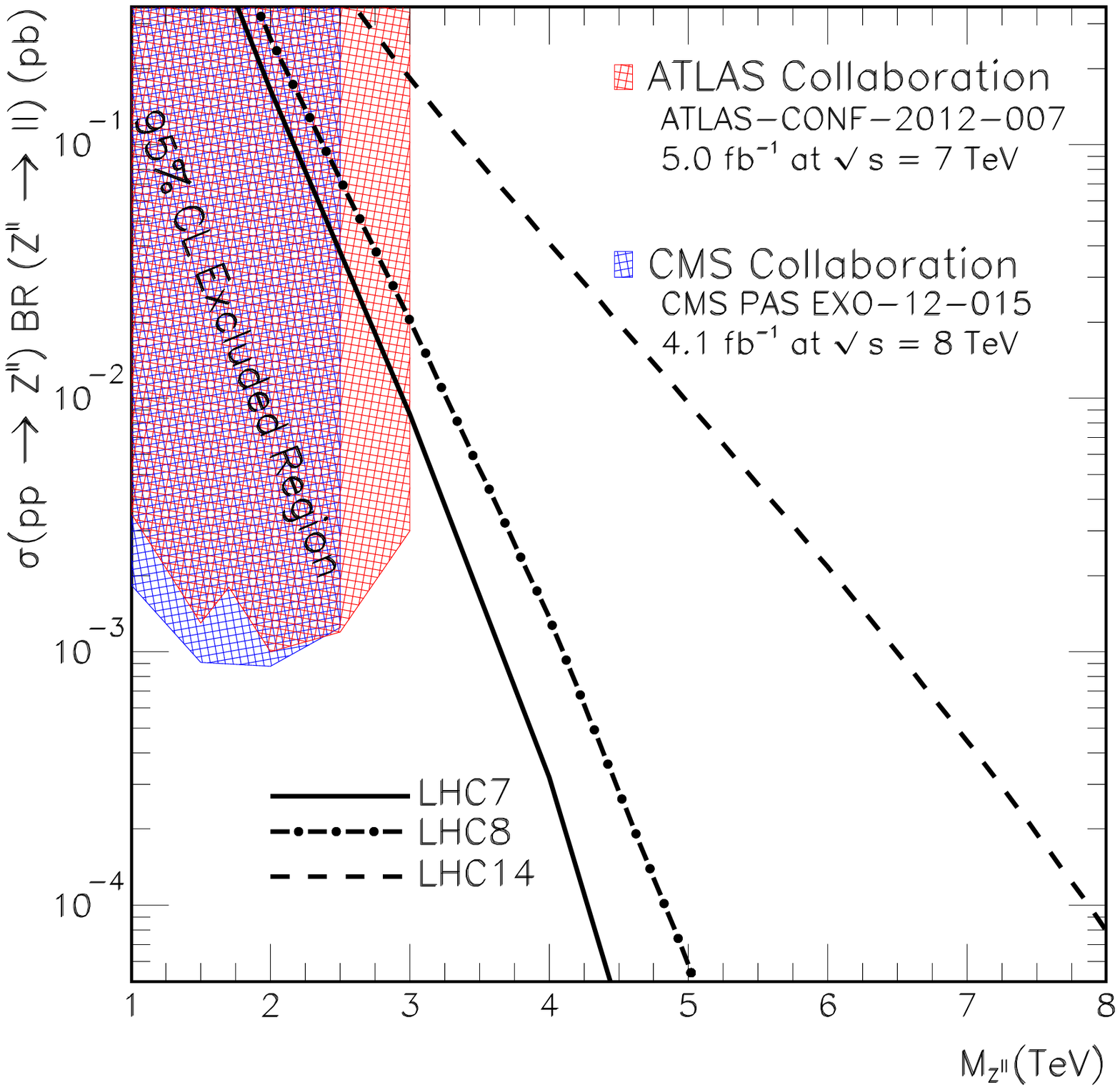}{0.99} \end{minipage}  
  \caption{Comparison of the (pre-cut) total cross section for the
    production of $p p \to Z'' \to \ell \ell$ with the 95\% CL upper
    limits on the production of a gauge boson decaying into two
    leptons, as reported by the ATLAS and CMS collaborations. The case
    in which $Z''$ is mostly diagonal in $I_R$ is shown in the left
    panel and the case in which it is mostly $B-L$ in the right
    panel.}
\label{fig:ATLAS} \end{figure}

The ATLAS Collaboration has searched for narrow resonances in the
invariant mass spectrum of dimuon and dielectron final states in event
samples at $\sqrt{s} = 7~{\rm TeV}$ corresponding to an integrated
luminosity of $4.9~{\rm fb}^{-1}$ and $5.0~{\rm fb}^{-1}$,
respectively~\cite{atlas_dilepton7tev}. The spectra are consistent
with SM expectations and thus upper limits on the cross section times
branching fraction for $Z''$ into lepton pairs have been set. More
recently, the CMS Collaboration updated the LHC7 results using $4.1~{\rm
  fb}^{-1}$ of data collected at $\sqrt{s} = 8~{\rm
  TeV}$~\cite{cms_dilepton8tev}. The combined upper limits from LHC7
and LHC8 are shown in Fig.~\ref{fig:ATLAS}.  Previous dilepton
searches by the LHC experiments have been reported
in~\cite{Collaboration:2011dca,Chatrchyan:2012it}.

In order to compare with these results we now turn to compute these
cross sections in our model.  The relevant part of (\ref{elef}), $f
\bar f Z''$ coupling, is of the form
\begin{eqnarray}
\mathscr {L} & = & \frac{1}{2}   \sqrt{g_Y^2 + g_2^2} \ \sum_f
\bigg(\epsilon_{f^i_L} \, \bar f_L^i \gamma^\mu f_L^i +
\epsilon_{f_R^i} \, \bar f_R^i \gamma^\mu f_R^i \bigg) \, Z''_\mu \, \nonumber \\
& = & \sum_f \bigg((g_{Y''}Q_{Y''})_{f_L^i} \, \bar f_L^i \gamma^\mu
f_L^i +  (g_{Y''}Q_{Y''})_{f_R^i} \, \bar f_R^i \gamma^\mu f_R^i \bigg) \, Z''_\mu \, ,
\label{lagrangian}
\end{eqnarray}
where $f_{L \, (R)}^i$ are fermion fields and $\epsilon_{f_L^i,f_R^i} = v_q \pm a_q$, with $v_q$ and $a_q$, the vector
and axial couplings respectively.  To compare our predictions with
LHC experimental searches in dilepton and dijets it is sufficient to
consider the production cross section in the narrow $Z''$ width
approximation,
\begin{equation}
\hat \sigma (q \bar q \to Z'')  =   K \frac{2 \pi}{3} \, \frac{G_F \, M_Z^2}{\sqrt{2}}  \left[v_q^2 (\phi, g'_1)+ a_q^2 (\phi, g'_1) \right] \, \delta \left(\hat s - M_{Z''}^2 \right) \,,
\end{equation}
where $G_F$ is the Fermi coupling constant and the $K$-factor
represents the enhancement from higher order QCD processes estimated
to be $K \simeq 1.3$~\cite{Barger}. After folding $\hat \sigma$ with
the CTEQ6 parton distribution functions~\cite{Pumplin:2002vw}, we
determine (at the parton level) the resonant production cross
section. In Figs.~\ref{fig:CMS} and \ref{fig:ATLAS} we compare the
predicted $\sigma (p p \to Z'') \times {\rm BR} (Z'' \to jj)$ and
$\sigma (p p \to Z'') \times {\rm BR} (Z'' \to \ell \ell)$ production
rates with 95\% CL upper limits recently reported by the CMS and ATLAS
collaborations. Selection cuts will probably reduce event rates by
factors of 20\%. Keeping this in mind, we conclude that if $Z''$ is
mostly $I_R$, then the predicted production rates for $M_{Z''} \approx
4~{\rm TeV}$ at $\sqrt{s} = 8~{\rm TeV}$ saturate the current dijet
limits.  On the other hand, if $Z''$ is mostly $B-L$ the lower limit
on the gauge boson mass, $M_{Z''} \agt 3~{\rm TeV}$, is determined
primarily from dilepton searches.

\begin{table}
  \caption{Chiral couplings of $Y$, $Z'$, and $Z''$ gauge bosons. All
    fields in a given set have a common $g_{Y'} Q_{Y'},\, g_{Y''}
    Q_{Y''}$ couplings. We have taken $Z'$ to be mostly $B$ and and $Z''$
    to be mostly $I_R$.}
\begin{tabular}{cccc}
\hline
\hline
~~~~~Fields~~~~~ & ~~~~~$g_Y Q_Y$~~~~~  & ~~~~~$g_{Y'} Q_{Y' }$~~~~~  &  ~~~~~$g_{Y''} Q_{Y''}$~~~~~  \\ \hline
$U_R$ &  $\phantom{-}0.2434$ & $\phantom{-}0.1836$ & $\phantom{-}0.3321$ \\
$D_R$ &  $-0.1214$ & $\phantom{-}0.1838$ & $-0.3933$ \\ 
$L_L$ &  $-0.1826$ & $\phantom{-}0.0759$ & $\phantom{-}0.0918$ \\ 
$E_R$ &   $-0.3650$ & $\phantom{-}0.0760$ & $-0.2709$ \\ 
$Q_L$ &  $\phantom{-}0.0610$ & $\phantom{-}0.1837$ & $-0.0306$ \\
$N_R$ &  $\phantom{-}0.0000$ & $\phantom{-}0.0758$ & $\phantom{-}0.4545$ \\
$H\phantom{''}$ &  $\phantom{-}0.1824$ & $\phantom{-}0.0000$ & $\phantom{-}0.3627$ \\
$H''$ &  $\phantom{-}0.0000$ & $-0.0758$ & $-0.4545$ \\
\hline
\hline
\end{tabular} 
\label{table-iv}
\end{table}

\begin{table}
  \caption{Chiral couplings of $Y$, $Z'$, and $Z''$ gauge bosons. All
    fields in a given set have a common $g_{Y'} Q_{Y'},\, g_{Y''}
    Q_{Y''}$ couplings.  We have taken $Z'$ to be mostly $L$ and and $Z''$
    to be mostly $B-L$.}
\begin{tabular}{cccc}
\hline
\hline
~~~~~Fields~~~~~ & ~~~~~$g_Y Q_Y$~~~~~  & ~~~~~$g_{Y'} Q_{Y' }$~~~~~  &  ~~~~~$g_{Y''} Q_{Y''}$~~~~~  \\ \hline
$U_i$ & $\phantom{-} 0.2435$ & $\phantom{-}0.1101$ & $-0.0763$ \\
$D_i$ &  $-0.1217$ & $\phantom{-}0.1101$ & $-0.2242$ \\ 
$L_i$ &  $-0.1825$ & $\phantom{-}0.7165$ & $\phantom{-}0.4509$ \\ 
$E_i$ & $-0.3651$ & $\phantom{-}0.7165$ & $\phantom{-}0.3769$ \\ 
$Q_i$ &  $\phantom{-}0.0609$ & $\phantom{-}0.1101$ & $-0.1503$ \\
$N_i$ &  $\phantom{-}0.0000$ & $\phantom{-}0.7165$ & $\phantom{-}0.5248$ \\
$H$ &  $\phantom{-} 0.1826$ & $-0.0000$ & $\phantom{-}0.0739$ \\
$H''$ &  $-0.0000$ & $-0.7165$ & $-0.5248$ \\
\hline
\hline
\end{tabular} 
\label{newtable}
\end{table}

For the discovery potential in the high mass region the dijet channel
is statistically a better discriminator than lepton pairs. Therefore,
we investigate (at the parton level) the LHC14 sensitivity for a
$Z''$ resonance (which is mostly $I_R$) in data binned according to the dijet invariant mass
$M$, after setting cuts on the different jet rapidities, $|y_1|, \,
|y_2| \le 1$ and transverse momenta $p_{\rm T}^{1,2}>50$~GeV.  With
the definitions $Y\equiv \frac{1}{2} (y_1 + y_2)$ and $y \equiv
\frac{1}{2} (y_1-y_2)$, the cross section per interval of $M$ for $p
p\rightarrow {\rm dijet}$ is given by
\begin{eqnarray}
\frac{d\sigma}{dM} & = & M\tau\ \sum_{ijkl}\left[
\int_{-Y_{\rm max}}^{0} dY \ f_i (x_a,\, M)  \right. \ f_j (x_b, \,M ) \
\int_{-(y_{\rm max} + Y)}^{y_{\rm max} + Y} dy
\left. \frac{d\sigma}{d\hat t}\right|_{ij\rightarrow kl}\ \frac{1}{\cosh^2
y} \nonumber \\
& + &\int_{0}^{Y_{\rm max}} dY \ f_i (x_a, \, M) \
f_j (x_b, M) \ \int_{-(y_{\rm max} - Y)}^{y_{\rm max} - Y} dy
\left. \left. \frac{d\sigma}{d\hat t}\right|_{ij\rightarrow kl}\
\frac{1}{\cosh^2 y} \right] \,,
\label{longBH}
\end{eqnarray}
where $f(x,M)$'s are parton distribution functions (we use  CTEQ6~\cite{Pumplin:2002vw}),
$\tau = M^2/s$, $x_a = \sqrt{\tau} e^{Y}$,  $x_b = \sqrt{\tau} e^{-Y},$ and
\begin{equation}
  |{\cal M}(ij \to kl) |^2 = 16 \pi \hat s^2 \,
  \left. \frac{d\sigma}{d\hat t} \right|_{ij \to kl} \, ;
\end{equation}
we specify partonic subprocesses with caret notation $(\hat s, \, \hat t, \, \hat u)$.
The $Y$ integration range in Eq.~(\ref{longBH}), $Y_{\rm max} = {\rm min} \{ \ln(1/\sqrt{\tau}),\ \ y_{\rm max}\}$, comes from requiring $x_a, \, x_b < 1$ together with the rapidity cuts $y_{\rm min} <|y_1|, \, |y_2| < y_{\rm max}$. The kinematics of the scattering also provides the relation $M = 2p_T \cosh y$, which when combined with $p_T = M/2 \ \sin \theta^* = M/2 \sqrt{1-\cos^2 \theta^*},$ yields $\cosh y = (1 - \cos^2 \theta^*)^{-1/2},$ where $\theta^*$ is the center-of-mass scattering angle.  Finally, the Mandelstam invariants occurring in the cross section are given by $\hat s = M^2,$ $\hat t = -\frac{1}{2} M^2\ e^{-y}/ \cosh y,$ and $\hat u = -\frac{1}{2} M^2\ e^{+y}/ \cosh y.$

The average square amplitude (for incoming quark $q$ and outgoing quark $q'$) is given by, \ba
  |{\cal M} (q\bar q \stackrel{Z''} {\to} q'\bar q {}')|^2  &= &\frac 1 4 \lsb g_{Y''}^2 Q_{Y''}^2(q_L)  + g_{Y''}^2 Q_{Y''}^2(q_R) \rsb \lsb g_{Y''}^2 Q_{Y''}^2(q_L{}')  +  g_{Y''}^2 Q_{Y''}^2(q_R{}') \rsb \nn
& \times & \left [\frac{2(  u^2+   t^2)}{( s-M_{Z''}^2)^2 + (\Gamma_{Z''}\ M_{Z''})^2} \right],
\label{Zdprimecross}
\ea where $g_{Y''}Q_{Y''}(q_L)$ and $g_{Y''} Q_{Y''}(q_R)$ are the
couplings of $Z''$ to quarks. (Note that we have not summed over the flavors,
but we did average and sum the colors).

The decay width of $Z'' \to f\bar f$ is given by
\begin{equation}
\Gamma( Z'' \to f \bar f) = \frac{G_F M_Z^2}{6 \pi \sqrt{2}}  N_c
 \, M_{Z''} \sqrt{1 -4x} \left[v_f^2 (1+2x) + a_f^2 (1-4x)
\right] \, \left(1 + \frac{\alpha_s}{\pi}\right) \, ,
\end{equation}
where  $\alpha_s = \alpha_s(M_{Z''})$ is the strong coupling constant at the
scale $M_{Z''}$, $x = m_f^2/M_{Z''}^2$, $v_f$ and $a_f$ are the vector
and axial couplings, and $N_c =3$ or 1 if $f$ is a quark or a lepton,
respectively~\cite{Feldman:2006wb}. For our fiducial values of $g'_1$ and $\phi$ we obtain
$v_u^2 + a_u^2 = 0.396$ and $v_d^2 + a_d^2 = 0.554$.

We calculate a signal-to-noise ratio, with the signal rate (${\cal
  S}$) estimated in the invariant mass window $[M_{Z''} - 2 \Gamma, \,
M_{Z''} + 2 \Gamma]$.  To accommodate the minimal acceptance cuts on
dijets from the CMS and ATLAS proposals~\cite{Bhatti:2008hz}, an
additional kinematic cut, $|y_{\rm max}|<1.0$, has been included in
the calculation. The noise (${\cal N}$) is defined as the square root
of the number of QCD background events (${\cal B}$) in the same dijet
mass interval for the same integrated luminosity. In Table.~\ref{table-v} we
show the behavior of the signal-to-noise ratio as a function of the
mass of $Z''$ at LHC14 for different integrated luminosities. We
conclude that the LHC provides a generous discovery potential for
$Z''$ which is mostly $I_R$. The discovery potential of a $Z''$ which
is mostly $B-L$ is controlled by the sensitivity of LHC14  to dilepton
final states. For $300~{\rm fb}^{-1}$, the projected sensitivity is 
$M_{Z''} \alt 5~{\rm TeV}$~\cite{Chiang:2011kq}.

\begin{table}[htb]
\caption{Signal-to-noise ratio at LHC14 for different integrated luminosities.}
\label{table-v}
\begin{tabular}{c|@{}ccc|@{}ccc|@{}ccc}
  \hline
  \hline
  & \multicolumn{3}{@{}c|}{10~fb$^{-1}$} & \multicolumn{3}{@{}c|}{100~fb$^{-1}$} & \multicolumn{3}{@{}c}{1000~fb$^{-1}$} \\
  \cline{2-4} \cline{5-7} \cline{8-10}
  ~~~$M_{Z''}~({\rm TeV})$~~~ & ~~~${\cal S}$~~~ & ~~~${\cal B}$~~~ & ~~~${\cal S}/{\cal N}$~~~ & ~~~${\cal S}$~~~ & ~~~${\cal B}$~~~ & ~~~${\cal S}/{\cal N}$~~~ & ~~~${\cal S}$~~~ & ~~~${\cal B}$~~~ & ~~~${\cal S}/{\cal N}$~~~ \\
  \hline
  4 & ~39 & ~579	& 1.62	  & 391   & 5789   & ~5.14   & ~3910  & ~57895     & 16.25 \\
  5 & ~~7& ~176	& 0.50	  & ~67   & 1759   & ~1.60   & ~~670  & ~17590      & ~5.05 \\
  6 & ~~1 & ~~66	& 0.14	  & ~11   & ~664   & ~0.44   & ~~113 & ~~6646      & ~1.39 \\
  \hline
  \hline
\end{tabular}
\end{table}

If the $Z''$ is observed at the LHC, we will obviously measure its
mass, its total width and cross section. In addition, the off- and on
resonance peak forward-backward charge asymmetries $A_{\rm FB}^\ell$
would provide additional information about $Z''$ couplings and
interference effects with the $Z$ boson and the
photon~\cite{Langacker:1984dc,Rosner:1995ft,Dittmar:1996my}.\footnote{The
  leptonic forward-backward charge asymmetry $A^{\ell}_{\rm FB}$ is
  defined from the lepton angular distribution with respect to the
  quark direction in the centre-of-mass frame  as $d \sigma/d\cos
  \theta^* \propto \frac{3}{8} ( 1 + \cos^2 \theta^*) + A_{\rm
    FB}^\ell \cos \theta^*$. The lepton angle $\theta^*$ in the
  dilepton center-of-mass frame can be calculated using the measured
  four momenta of the dilepton system; $A_{\rm FB}^\ell$ can then be
  determined with an unbinned maximum likelihood fit to the $\cos
  \theta^*$ distribution.}  Besides, the $Z''$ rapidity distribution
is sensitive to the $u \bar u Z''$ and $b \bar b Z''$ gauge couplings.
Since the $W^\pm$ and $Z$ boson rapidity distributions will be
measured in great detail at the LHC, rapidity spectra for the mass
region of interest can be calculated separately for $u \bar u$, $d
\bar d$, and sea quark antiquark annihilation. A combined fit to the
relative parton distribution functions and the $Z''$ rapidity
distribution would allow us to obtain the fraction of $Z''$ bosons
produced from $u \bar u$ and $d \bar d$ initial
states~\cite{Dittmar:2003ir}.

\section{Neutrino Cosmology Redux}
\label{SEC-IV}

In this section we reexamine some critical cosmological issues
surrounding the presence of the six additional neutrino degrees of
freedom correlated to the presence of $Z''$ in our dynamical D-brane
model. These considerations, when viewed in the context of most recent
cosmological observations are found to constrain the mass of the $Z''$
to an interesting narrow band, which will be directly probed by
LHC14. To provide a starting point, we first summarize the
``best-fit''  cosmological parameters to recent data.

\subsection{Beyond $\bm{\Lambda}$CDM}

Our universe seems, according to the present-day evidence, to be
spatially flat and to posses a non-vanishing cosmological constant
($\Lambda$) plus cold dark matter (CDM), corresponding respectively to
roughly 70\% and 25\% of the total density, with the remaining 5\% in
baryons.  The standard $\Lambda$CDM cosmology provides a rather good
fit of existing data from BBN ($\sim~20~{\rm minutes}$), the CMB
($\sim~380~{\rm kyr}$), and the galaxy formation epochs of the
universe ($\agt~1~{\rm Gyr}$). However, there are also tantalizing
hints for the presence of an extra relativistic component, dubbed dark
radiation.

Taking these hints at face value, the most straightforward variation
of standard $\Lambda$CDM is ``extra'' energy contributed by new
relativistic particles ``$X$.''  When the $X$'s don't share in the
energy released by $e^+e^-$ annihilation, it is convenient to account
for the extra contribution to the SM energy density, by normalizing it
to that of an ``equivalent'' neutrino species~\cite{Steigman:1977kc}
\begin{equation} 
\rho_{X} \equiv \Delta N_{\nu} \, \rho_{\nu} = 
{7 \over 8}\Delta N_{\nu} \, \rho_{\gamma}, 
\label{deln} 
\end{equation} 
where $\rho_\nu$ is the energy density in neutrinos and $\rho_\gamma$
is the energy density in photons (which by today have redshifted to
become the CMB photons at a temperature of about
2.7~K).  For each additional ``neutrino-like''
particle ({\em i.e.} any two-component fermion), if $T_{X} = T_{\nu_L}$,
then $\Delta N_\nu = 1$; if $X$ is a scalar (and $T_{X} = T_{\nu_L}$),
then $\Delta N_\nu = 4/7$.  However, it may well be that the $X$'s
have decoupled even earlier in the evolution of the universe and have
failed to profit from the heating when various other
particle-antiparticle pairs annihilated (or unstable particles
decayed).  In this case, the contribution to $\Delta N_\nu$ from each
such particle will be $< 1$ and $< 4/7$, respectively.  The
contribution of the $2.984 \pm 0.009$ neutrino species (measured from
the width for invisible $\nu \bar \nu$ decays of the $Z$
boson~\cite{:2005ema}) to $N_\nu^{\rm eff} = N_\nu + \Delta N_\nu$ is
$N_\nu = 3.046$; the small deviation from 3 is due to partial heating
of neutrinos in the early universe by $e^+ e^-$ annihilation, see {\em
  e.g.}~\cite{Mangano:2005cc}.

The competition between gravitational potential and pressure gradients
is responsible for the peaks and troughs in the CMB temperature
angular power spectrum, see {\it e.g.}~\cite{Dodelson:2003ft}. The
redshift $z_{\rm eq}$ of matter-radiation equality, \be 1 + z_{\rm
  eq}= \frac{\Omega_{\rm m} h^2}{\Omega_{\rm R} h^2} =
\frac{\Omega_{\rm m} h^2}{\Omega_\gamma h^2} \left[1 + \frac{7}{8}
  \left(\frac{4}{11}\right)^{4/3} N_\nu^{\rm eff} \right]^{-1} \,,\ee
affects the time (redshift) duration over which this competition
occurs.  Here, $\Omega_{\rm m} h^2$ is the total matter density
(comprised, for nearly massless neutrinos, of baryons and CDM), $h$
($H_0 \equiv 100h~{\rm km/s/Mpc}$) is the normalized Hubble constant,
and $\Omega_\gamma h^2 = 2.469 \times 10^{-5}$ is the present-day
photon energy density.  The primary effect of extra relativistic
degrees of freedom on the CMB results essentially from changing the
redshift of matter-radiation equality.  If the radiation content is
increased, matter-radiation equality is delayed, and occurs closer (in
time and/or redshift) to the epoch of recombination.  This implies the
universe is younger at recombination with a correspondingly smaller
sound horizon $s_*$.  Since the location of the $n^{\rm th}$ peak in
the angular power spectrum scales roughly as $n\pi D_*/s_*$ (where
$D_*$ is the comoving angular diameter distance to
recombination\footnote{The angles on the sky are related to actual
  physical distance via the angular diameter distance $d$, defined as
  the ratio of the physical length (transverse to the line of sight)
  and the angle it covers $d \equiv \lambda_{\rm
    phys}/\vartheta$. Likewise, $D \equiv \lambda^c/\vartheta$, where
  $\lambda^c = (1+ z) \lambda_{\rm phys}$ is the corresponding
  comoving length and $z$ the redhsift; $D = (1+z) d$.}), if $\Delta
N_\nu > 0$ the peaks shift to smaller angular scales and with greater
separation~\cite{Dodelson:2003ft}. Therefore, the equality redshift is one of the fundamental
observables that one can extract from WMAP data, mainly from the
height of the third acoustic peak relative to the first peak.

The variation in
$N_\nu^{\rm eff}$ reads~\cite{Komatsu:2010fb}
\begin{equation}
\frac{\delta N_\nu^{\rm eff}}{N_\nu^{\rm eff}} \simeq 2.45 \ \frac{\delta(\Omega_m h^2)}{\Omega_m h^2}
 - 2.45 \ \frac{\delta z_{\rm eq}} { 1 + z_{\rm eq}} \, .
\end{equation}
The latest distance measurements from the Baryon Acoustic Oscillations
(BAO) in the distribution of galaxies~\cite{Percival:2009xn} and
precise measurements of the Hubble constant $H_0$~\cite{Riess:2009pu}
provide an independent determination in the fractional error in
$\Omega_m h^2$ and allow a precise determination of $N_\nu^{\rm eff}$.
The parameter constraints from the combination of WMAP 7-year data,
BAO, and $H_0$ lead to $N_\nu^{\rm eff} =
4.34^{+0.86}_{-0.88}$~\cite{Komatsu:2010fb}.  Similarly, a combination
of BAO and $H_0$ with data from the Atacama Cosmology Telescope (ACT)
yields $N_\nu^{\rm eff} = 4.6 \pm 0.8$~\cite{Dunkley:2010ge}, whereas
data collected with the South Pole Telescope (SPT) combined with BAO
and $H_0$ arrive at $N_\nu^{\rm eff} = 3.86 \pm
0.42$~\cite{Keisler:2011aw}. Although none of these measurements
individually deviates from the standard value by more than about two
standard deviations, they collectively rule out $N_\nu = 3.046$ at the
approximately 99\% CL, and instead prefer roughly one
extra effective neutrinos species~\cite{Hamann:2011hu}.\footnote{A
  more recent study seems to indicate $3.0 < N_\nu^{\rm eff} < 4.1$~\cite{Moresco:2012by}.}

The expansion rate of the universe at early times increases with the
number of relativistic particle species in thermal equilibrium, and
this in turn sets timescales for
BBN~\cite{Sarkar:1995dd,Olive:1999ij,Steigman:2007xt}. One can then
use the BBN yields of light nuclei to constrain the number of light
species quantitatively. The nucleosynthesis chain begins with the
formation of deuterium in the process $p(n,\gamma){\rm D}$. However,
photo-dissociation by the high number density of photons delays
production of deuterium (and other complex nuclei) until well after
$T$ drops below the binding energy of deuterium, $\Delta_{\rm D} =
2.23~{\rm MeV}$. The number of photons per baryon above the deuterium
photo-dissociation threshold, $\eta^{-1} e^{-\Delta_{\rm D}/T}$, falls
below unity at $T \simeq 0.1~{\rm MeV}$, where $\eta \equiv
n_B/n_\gamma \sim 5 \times 10^{-10}$ is the baryon to photon number
density. Nuclei can then begin to form without being immediately
photo-dissociated again. Only 2-body reactions such as ${\rm
  D}(p,\gamma)^3{\rm He},$ $^3{\rm He}({\rm D},p)^4{\rm He},$ are
important because the density is rather low at this time. Nearly all
the surviving neutrons when nucleosynthesis begins end up bound in the
most stable light element $^{4}{\rm He}$. Heavier nuclei do not form
in any significant quantity both because of the absence of stable
nuclei with mass number 5 or 8 (which impedes nucleosynthesis via
$n^4{\rm He}$, $p^4{\rm He}$, or $^4{\rm He}^4{\rm He}$ reactions) and
the large Coulomb barriers for reactions such as ${\rm T}(^4{\rm
  He},\gamma)^7{\rm Li}$ and $^3{\rm He}(^4{\rm He},\gamma)^7{\rm
  Be}$. Hence the primordial mass fraction of $^4{\rm He}$,
conventionally referred to as $Y_{\rm p}$, can be estimated by the
simple counting argument
\begin{equation}
Y_{\rm p} = \frac{2(n/p)}{1 + n/p} \  .
\end{equation}
For $T \agt 1~{\rm MeV}$, weak interactions were in thermal
equilibrium, thus fixing the ratio of the neutron and proton number
densities to be $n/p = e^{-Q/T}$, where $Q = 1.293~{\rm MeV}$ is the
neutron-proton mass difference. As the temperature dropped, the
neutron-proton inter-conversion rate, $\Gamma_{n \leftrightharpoons p}
\sim G_F^2 T^5$, fell faster than the Hubble expansion rate, $H \approx
\sqrt{N(T)} \ T^2/M_{\rm Pl}$ (see {\it e.g.}~\cite{Kolb:1990vq}). Since $N(T)$ counts the number of relativistic
particle species determining the energy density in radiation, the
neutron fraction $n/p$ is directly sensitive to $\Delta N_\nu$. For
standard $\Lambda$CDM, the  freeze-out temperature of $\nu_L$ is 
\begin{equation}
T_{\rm FO} \sim \left[\frac{\sqrt{N(T_{\rm FO})}}{M_{\rm Pl} \ 
    G_F^2}\right]^{1/3} \simeq 1~{\rm MeV} \,, 
\end{equation}
yielding $n/p \simeq 1/6$ and $Y_{\rm p} \simeq 0.25$.

The evidence for extra radiation from $Y_{\rm p}$ data is, however,  somewhat
ambiguous. The observationally-inferred primordial fractions of
baryonic mass in $^{4}$He ($Y_{\rm p} = 0.2472 \pm
0.0012$~\cite{Izotov:2007ed}, $Y_{\rm p} = 0.2477 \pm
0.0029$~\cite{Peimbert:2007vm}, and $Y_{\rm p} = 0.250 \pm
0.004$~\cite{Fukugita:2006xy}) have been constantly favoring
$N_\nu^{\rm eff} \alt 3$~\cite{Simha:2008zj}.  Unexpectedly, two
recent independent studies determined $Y_{\rm p} = 0.2565 \pm 0.001
({\rm stat}) \pm 0.005 ({\rm syst})$~\cite{Izotov:2010ca} and $Y_{\rm
  p} = 0.2561 \pm 0.0108$~\cite{Aver:2010wq}. For $\tau_n = 885.4 \pm 0.9~{\rm s}$ and
$\tau_n = 878.5 \pm 0.8~{\rm s}$, the updated effective number of
light neutrino species is reported as $N_\nu^{\rm eff} =
3.68^{+0.80}_{-0.70}$ ($2\sigma$) and $N_\nu^{\rm eff} =
3.80^{+0.80}_{-0.70}$ ($2\sigma$), respectively~\cite{Izotov:2010ca}.\footnote{For several years the Particle Data
  Group recommended $\tau_n = 885.7 \pm 0.8~{\rm
    s}$~\cite{Hagiwara:2002fs}.  More recently, conflicting lifetimes
  $\tau_n = 878.5 \pm 0.7 \pm 0.3~{\rm s}$~\cite{Serebrov:2007ve} and
  $\tau_n = 880.7 \pm 1.3 \pm 1.2~{\rm s}$~\cite{Pichlmaier:2010zz}
  have been reported. The Particle Data Group now recommends a world
  average that includes the conflicting values, $\tau_n = 881.5 \pm
  1.5~{\rm s}$~\cite{Nakamura:2010zzi}, with errors that have been
  inflated to reflect the discrepancy.}

As mentioned above, the primordial deuterium abundance depends not
just on $N_\nu^{\rm eff}$ but also on the cosmological baryon density,
$\Omega_b h^2$. Prior to the precise inference of $\Omega_b h^2$ from
CMB measurements, the strongest constraint on $N_\nu^{\rm eff}$ came
from using the deuterium-to-hydrogen number ratio D/H to restrict
$\Omega_b h^2$ and then exploiting the $N_\nu^{\rm eff}$ dependence of
the primordial helium mass fraction $Y_{\rm p}$. However, D/H has its
own dependence on $N_\nu^{\rm
  eff}$~\cite{Cardall:1996ec,Steigman:2010pa}; a strong external
constraint on $ \Omega_b h^2$ allows BBN constraints on $N_\nu^{\rm
  eff}$ that are independent of $Y_{\rm p}$. Since precise
measurements of $Y_{\rm p}$ are difficult, the constraint on
$N_\nu^{\rm eff}$ from D/H is found to be competitive with that from
$Y_{\rm p}$. A recent analysis, which combines the CMB results with
BBN theory and the observed D/H, suggests $N_\nu^{\rm eff} = 3.90 \pm
0.44$~\cite{Nollett:2011aa}.\footnote{The BBN calculations in this
  analysis includes updates of nuclear rates in light of recent
  experimental and theoretical information, with the most significant
  change occurring for the $d(p,\gamma)^3{\rm He}$ cross section.}

In summary, though uncertainties remain large, the most recent
cosmological observations show a consistent preference for additional
relativistic degrees of freedom (r.d.o.f.) during BBN and the CMB
epochs. We take these hints as motivation for the subsequent analysis,
which consists of the following tasks: {\em (1)} to explain the dark
radiation using the non-supersymmetric $U(3)_B \times Sp(1)_L \times
U(1)_L \times U(1)_{I_R}$ D-brane model, in
which the additional r.d.o.f. are the three flavors of light
right-handed neutrinos which interact with the SM fermions via the
exchange of heavy vector fields $Z'$ and $Z''$; {\em (2)} to suppress
the six additional fermionic r.d.o.f. to levels in compliance with BBN
and CMB. This is accomplished by imposing the decoupling of $\nu_R$'s
from the plasma {\em early enough} so that they undergo incomplete
reheating during the quark-hadron transition; and {\em late enough} so
as to leave an excess neutrino density suggested by the data. These
requirements strongly constrain the masses of the heavy vector
fields. Together with the couplings given in Table~\ref{table-iv}, the
model is fully predictive, and can be confronted with dijet and
dilepton data  from LHC8 and, eventually, LHC14.

\subsection{Cosmology of Intersecting Branes}

The ensuing discussion will be framed in the context of a $Z''$ which
is mostly $I_R$, and we will comment on the case in which $Z''$ is
mostly $B-L$ after presenting our results.

We begin by first establishing, in a model independent manner, the
range of decoupling temperatures implied by the  BBN and CMB
analyses. For the subsequent study, the physics of
interest will be taking place at energies in the region of the
quark-hadron transition, so that we will restrict ourselves to the
following fermionic fields, and their contribution to r.d.o.f.: $
\left[ 3 u_R \right] + \left[ 3 d_R \right] + \left[ 3s_R\right ] +
\left[ 3 \nu_L + e_L + \mu_L \right] + \left[ e_R + \mu_R \right]+
\left[ 3 u_L + 3d_L + 3s_L\right]+ \left[ 3 \nu_R \right]$. This
amounts to 28 Weyl fields, translating to 56 fermionic r.d.o.f.

Next, in line with our stated plan, we use the data estimate to
calculate the range of decoupling temperature. The effective number of
neutrino species contributing to r.d.o.f. can be written as $
N_\nu^{\rm eff} = 3 [1 + (T_{\nu_R}/T_{\nu_L}) ^4] \,;$ therefore,
taking into account the isentropic heating of the rest of the plasma
between $\nu_R$ decoupling temperature $T_{\rm dec}$ and the end of
the reheating phase,
\begin{equation}
\Delta N_\nu = 3 \left(\frac{N(T_{\rm end})}{N(T_{\rm dec})} \right)^{4/3} \,,
\label{7}
\end{equation}
where $T_{\rm end}$ is the temperature at the end of the reheating
phase, and $N(T) = r(T)(N_{\rm B} + \frac{7}{8} N_{\rm F})$ is the
effective number of r.d.o.f.  at temperature $T$, with $N_{\rm B} = 2$
for each real vector field and $N_{\rm F} = 2$ for each
spin-$\frac{1}{2}$ Weyl field. The coefficient $r(T)$ is unity for the
lepton and photon contributions, and is the ratio $s(T)/s_{\rm SB}$
for the quark-gluon plasma. Here $s(T) (s_{\rm SB})$ is the actual
(ideal Stefan-Bolzmann) entropy. Hence $N(T_{\rm dec}) = 47.5~r(T_{\rm
  dec}) + 14.25.$ We take $N(T_{\rm end}) = 10.75$ reflecting
$(e_L^-+e_R^+ + e_R^- + e_L^+ \nu_{eL} + \bar \nu_{eR} + \nu_{\mu L} +
\bar \nu_{\mu R} + \nu_{\tau L} + \bar \nu_{\tau R} + \gamma_L +
\gamma_R)$. We consistently omit $\nu_R$ in considering the
thermodynamics part of the discussion, but will include it when
dealing with expansion.  As stated in the introduction
\begin{equation}
\Delta N_\nu = \left \{ \begin{array}{ccl} 0.68^{+0.40}_{-0.35} & ~~(1
    \sigma) & ~~{\rm BBN} + Y_{\rm p} \\
0.90^{+0.44}_{-0.44} & ~~(1\sigma) & ~~{\rm CMB} + {\rm BBN} + {\rm H/D}
\end{array} \right.
\label{8}
\end{equation} 
so the excess r.d.o.f.  will lie within $1 \sigma$ of the central
value of each set of observations if $0.46 < \Delta N_\nu <
1.08$. From Eqs.~(\ref{7}) and (\ref{8}), the allowable range for $N$
is $23 < N (T_{\rm dec} )< 44.$ This is achieved for $0.18 < r(T_{\rm
  dec}) < 0.63.$ By comparing to Fig.~8 in Ref.~\cite{Bazavov:2009zn},
this can be translated into a temperature range
\begin{equation}
175~{\rm MeV} < T_{\rm dec} < 250~{\rm MeV} \, ,
\label{10}
\end{equation}
with the lower temperature coinciding with the region of most rapid
rise of the entropy.  Thus, the data implies that the $\nu_R$
decoupling takes place during the quark-hadron transition.

We now turn to use our model in conjunction with the decoupling condition to constrain its parameters. To this end we calculate the interaction rate
$\Gamma(T)$ for a right-handed neutrino and determine $T_{\rm dec}$ from the plasma via the prescription
\begin{equation}
 \Gamma(T_{\rm dec}) = H(T_{\rm dec}) \, .
\label{haim1}
\end{equation}

Let $f_L^i$ be a single species of Weyl fermion, representing the two r.d.o.f. 
$\{f_L^i, \bar f_R^i \}$, where the superscript indicates  bins $i = 3,5$. Similarly 
$f_R^i \in  \{f_R^i, \bar f_L^i\},$ for $i = 1,2,4,6$. Notice that the subscripts $L,R$ 
denote the actual helicities of the massless particles in question, not the chirality of the fields. With this 
said, we may write the amplitude for $f_L^i$ scattering
\begin{equation}
{\cal M} \left(\nu_R(p_1) f_L^i (p_2) \to \nu_R (p_3) f_L^i(p_4)\right) 
= \frac{G_i}{\sqrt{2}} [\bar u(p_3)  \gamma^\mu  (1+\gamma_5)  u(p_1)] [\bar u(p_4)  \gamma_\mu  
(1- \gamma_5)  u(p_2) ] \, .
\end{equation}
The other 3 amplitudes are obtained by the crossing substitutions in the second square bracket; for scattering from
\begin{eqnarray}
\bar f_R^i & \to & \bar v(p_2) \, \gamma_\mu \, (1- \gamma_5) \, v(p_4) \nonumber \\
f_R^i & \to & \bar u(p_4) \, \gamma_\mu \, (1+\gamma_5) \, u(p_2) \\
\bar f_L^i & \to & \bar v (p_2) \,  \gamma_\mu \, (1+\gamma_5) \, v (p_4) \, . \nonumber 
\end{eqnarray}
The cross sections for the  four  scattering processes (no average over helicities) are
\begin{equation}
\sigma\left(\nu_R f_L^i \to \nu_R f_L^i \right) =\frac{1}{3} \sigma\left(\nu_R \bar f_R^i \to \nu_R \bar f_R^i \right) = \frac{2}{3} \frac{G_i^2 s}{\pi} \quad ({\rm for \, bins} \, 
i= 
3,5)
\end{equation}
 and
\begin{equation}
\sigma\left(\nu_R \bar f_L^i \to \nu_R \bar f_L^i \right) = \frac{1}{3} \sigma(\nu_R f_R^i \to \nu_R f_R^i) = 
\frac{2}{3} \frac{G_i^2 s}{\pi} \quad ({\rm for \, bins}\, i = 1,2,4,6)  \, .
\end{equation}
In addition to these scattering processes, the $\nu_R$ interacts with the plasma through the annihilation processes: $\nu_R \bar\nu_L \to f_L^i \bar f_R^i,$ for  bins $i 
=3,5$, and
$\nu_R \bar\nu_L \to \, f_R^i \bar f_L^i$, for bins $i= 1,2,4,6.$ These all yield cross sections
$2 G_i^2 s/ (3\pi)$ due to  forward and backward suppression. Assuming all chemical potentials to be zero, the plasma will have an equal number density $n(T) = 
0.0913 
T^3$, for each fermion r.d.o.f. Thus,
\begin{equation}
\Gamma^{\rm scat} (T) = n(T) \left\langle \sum_{i=1}^6 \sigma_i(s) \, v_M \, {\cal N}_i \right \rangle \,, 
\end{equation}
where
$v_{\rm M} = 1- \cos \theta_{12}$ is the M\/{o}ller velocity, $s = 2 k_1 k_2 ( 1- \cos \theta_{12})$ is the square of the center-of-mass energy, and ${\cal N}_i$ is the 
multiplicty of Weyl fields in each bin (e.g., for $i =3, \,  {\cal N}_3 = 3 +2 = 5$). The scattering cross section is given by
\begin{equation}
\sigma_i^{\rm scat} = \sigma(\nu_R f_L^i \to \nu_R f_L^i) + \sigma(\nu_R \bar f_R^i \to \nu_R \bar f_R^i) = \frac{4}{3} \frac{2 G_i^2 \, s}{\pi} \quad {\rm for \, each} \, i = 
1, 
\dots, 6 \, ;
\label{15}
\end{equation}
similarly,
\begin{equation}
\sigma_i^{\rm ann} (s) = \sigma(\nu_R \bar \nu_L \to f_L^i \bar f_R^i + f_R^i \bar f_L^i) = \frac{1}{3} \frac{2 G_i^2 s}{\pi} \quad {\rm for \, each} \, i = 1, \dots 6 \, .
\label{16}
\end{equation}
Since $s = 2 k_1 k_2 (1 - \cos \theta_{12})$ and $v_M = 1 - \cos\theta_{12}$, we perform an approximate angular average $\langle (1- \cos\theta_{12})^2 \rangle = 4/3$, 
followed by a thermal averaging $\langle 2 k_1 k_2\rangle = 2 (3.15^2 \, T^2)$ to give
\begin{equation}
\Gamma^{\rm scat} (T)  = \left(\frac{4}{3}\right)^2 \, \frac{2}{\pi} \ 2  \ (3.15 T)^2 \, (0.0919 T^3)  \, \underbrace{\left(\sum_{i=1}^6 G_i^2 {\cal N}_i \right)}_{G_{\rm eff}^2} 
\simeq 2.05 G_{\rm eff}^2 \, T^5 \, .
\label{17}
\end{equation}
>From (\ref{15}), (\ref{16}), and (\ref{17}),
\begin{equation}
 \Gamma^{\rm ann}(T)  = \frac{1}{4}  \Gamma^{\rm scat} (T)  \simeq 0.50 \ G_{\rm eff}^2 \ T^5 \, .
\end{equation}
Each of the $G_i$ is given by the sum of the contributions from $Z'$ and $Z''$ exchange,
\begin{equation}
4 \frac{G_i}{\sqrt{2}} = \frac{g'_6 \ g'_i}{M^2_{Z'}} + \frac{g''_6 \ g''_i}{M^2_{Z''}} \, .
\end{equation}

%%%%%%%%%%%%%%%%%%%%%%%
\begin{figure}[tbp]
\postscript{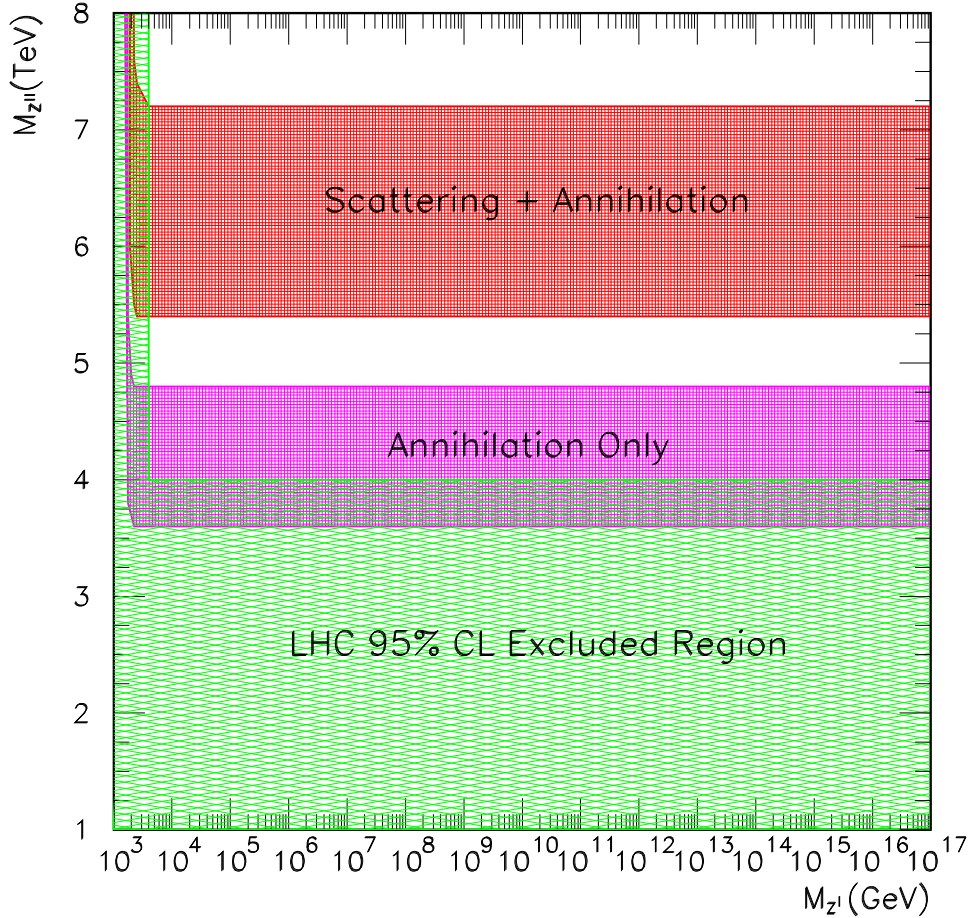}{0.80}
\caption{The shaded areas show the region allowed from decoupling
  requirements to accommodate CMB and BBN data. The hatched region
  indicates the masses excluded by the LHC8 dijet searches. The lower
  and upper shaded areas  pertain to  chemical and thermal 
 equilibrium, respectively.  These two estimates should serve
  to bracket the size of the actual effect.}
\label{fig:rnus}
\end{figure}
%%%%%%%%%%%%%%%%%%%%%%%%%%

The Hubble expansion parameter during this time is
\begin{equation}
H(T) = 1.66 \ \langle N(T) \rangle^{1/2} \ T^2/M_{\rm Pl} \,,
\label{haim2}
\end{equation}
where $M_{\rm Pl}$ is the Planck mass. Since the quark-gluon energy
density in the plasma has a similar $T$ dependence to that of the
entropy (see Fig.~7 in~\cite{Bazavov:2009zn}), we take $N(T) =
47.5~r(T) + 19.5,$ so that $H(T) = 10.3~T^2/M_{\rm Pl}$. (The first
factor provides an average for $r(T)$ over the temperature region, and
we have now included the six $\nu_R$ r.d.o.f.) Since $\Gamma \propto
T^5$ and $H \sim T^2$, it is clear that if at some temperature $T_{\rm
  dec}$, $H(T_{\rm dec}) = \Gamma_i (T_{\rm dec})$, the ratio
$\Gamma/H$ will fall rapidly on further cooling. Thus from (\ref
{haim1}) and (\ref{haim2}) the equation determining $T_{\rm dec}$
depends on: {\em (1)} whether we need to preserve the absence of a
chemical potential, or {\em (2)} whether we need simply to mantain
physical equilibrium. The decoupling condition in these two cases is:
{\em (1)} $\Gamma^{\rm ann} (T_{\rm dec}) = H (T_{\rm dec})$ and {\em
  (2)} $\Gamma^{\rm scat} (T_{\rm dec})+ \Gamma^{\rm ann} (T_{\rm
  dec}) = H (T_{\rm dec})$; or numerically: {\em (1)}
\begin{equation}
0.50~G_{\rm eff}^2~T^5_{\rm dec} = 10.3~T_{\rm dec}^2/M_{\rm pl} \Rightarrow T^3_{\rm dec} = 20.6~(G^2_{\rm eff} M_{\rm Pl})^{-1} \,,
\end{equation}
and {\em (2)} 
\begin{equation}
2.50~G_{\rm eff}^2 \, T_{\rm dec}^5 = 10.3~T_{\rm dec}^2/M_{\rm Pl} \Rightarrow T_{\rm dec}^3 = 4.1~(G^2_{\rm eff} M_{\rm Pl})^{-1} \, .
\end{equation}
$T_{\rm dec}$ as determined from these equations must lie in the band
(\ref{10}).  

Since all freedom of determining coupling constant and mixing angles
has been exercised, there remains only constraints on the possible
values of $M_{Z'}$ and $M_{Z''}$. For high mass string scales the
contribution from $M_{Z'}$ to $G_{\rm eff}$ is neglible. We find that
for certain ranges of $M_{Z''}$ the decoupling of the $\nu_R$'s occurs
during the course of the quark-hadron transition, just so that they
are only partially reheated compared to the $\nu_L$'s --- the desired
outcome.  Since our aim is to match the data, which has lower and
upper bounds on the neutrino ``excess'', we obtain corresponding upper
and lower bounds on the $Z''$ gauge field mass. Roughly speaking, if
decoupling requires a freezout of the annihilation channel (loss of
chemical equilibrium), then $3.6~{\rm TeV} < M_{Z''}< 4.8~{\rm
  TeV}$. This range will be probed at LHC14.  If thermal
equilibrium via scattering is sufficient, then $5.4~{\rm TeV} <
M_{Z''}< 7.4~{\rm TeV}$. 

Depending on the details of the string type model and $M_s$ some of
the couplings may go up and some may go down, but the net result for
$G_{\rm eff}$ involving the product of all these couplings is
virtually unchanged. Moreover, we have verified that if $M_s$ is
pushed downwards to the TeV-scale region both $M_{Z'}$ and $M_{Z''}$
contribute to $G_{\rm eff}$ and are within the LHC reach. A summary of
LHC7 constraints and $M_{Z'}-M_{Z''}$ mass regions consistent with CMB
+ BBN + $Y_{\rm p}$ + H/D data (within $1\sigma$) is encapsulated in
Fig.~\ref{fig:rnus}.

We comment briefly on the case in which $Z''$ is mostly $B-L$. By
comparing Tables~\ref{table-iv} and \ref{newtable} it becomes evident
that the $Z''$ coupling to neutrinos is stronger when the extra gauge
boson is almost diagonal in $B-L$. As a consequence, the allowed range
of masses from decoupling requirements to accommodate CMB and BBN data
is shifted to higher values: $4.5~{\rm TeV} < M_{Z''}< 6.1~{\rm TeV}$
if decoupling requires a freezout of the annihilation channel, and
$6.3~{\rm TeV} < M_{Z''}< 8.2~{\rm TeV}$ if thermal equilibrium via
scattering is sufficient.

The first cosmology results from the Planck satellite anticipated in
early 2013 would allow determination of $N_\nu^{\rm eff}$ with a
standard deviation of about 0.3, whereas the future Large Synoptic
Survey Telescope (LSST) could determine $N_\nu^{\rm eff}$ with a
standard deviation of about 0.1~\cite{Joudaki:2011nw}. These
observations when combined with future LHC results can directly test
the viability of our model.

\section{Supersymmetric Extension}
\label{SEC-V}

When the string scale is at high energies, supersymmetry is in
principle welcome for the hierarchy problem. Gauge bosons of the brane
stacks belong then to ${\cal N}=1$ vector multiplets together with the
corresponding gauginos, while at brane intersections chiral fermions
belong to chiral multiplets denoted by their left-handed fermionic
components $Q, L, U^c, D^c, E^c, N^c$, where the superscript $c$
stands for the charged conjugate in the familiar notation.  Moreover,
in the $(P,R)$ intersection, one should have the two usual Higgs
doublets chiral multiplets $H_1,H_2$ with the quantum numbers of $H^*$
and $H$, respectively. Finally, the extra Higgs singlet $H''$ becomes
naturally the superpartner of the right-neutrino superfield $N^c$.
The Yukawa interactions (\ref{Yukawas}) are now replaced by the
superpotential: \be W_Y=Y_u \, Q \, H_2 \, U^c+Y_d \, Q \, H_1 \, D^c
+Y_e \, L \, H_1 \, E^c+ Y_N \, L \, H_2 \, N^c \, .  \ee On
electroweak symmetry breaking, $H_2$ develops a VEV, as a result of
which $N^c$ couples with $\nu_L$ to form a Dirac neutrino. Since
superpotentials such as $M N^c N^c$ or $SN^cN^c$ are precluded by the
$U(1)_L$ and $U(1)_{I_R}$ gauge invariances, there seems no equivalent
of the seesaw mechanism to generate the Weinberg
term~\cite{Weinberg:1980bf} which gives rise to Majorana
neutrinos.\footnote{However it is possible that D-brane instantons
    can generate Majorana masses for these perturbatively forbidden
    operators~\cite{Blumenhagen:2006xt,Ibanez:2006da}.} Here $M$ is a
Majorana mass matrix in flavor space and $S$ is a gauge singlet. In
addition, the existence of the VEV $\langle N^c \rangle$ breaks the
$U(1)_L$ lepton gauge symmetry which allows the $Z''$ to grow a mass.
It also generates the $R$-parity breaking term $LH_2$, whose
coefficient is subject to a variety of phenomenological
constraints~\cite{Hall:1983id}.

A superfield $H''$ with $I_R = L = +1$ opposite to $N^c$ presents
difficulties. A VEV for this version of $H''$ serves equally well for
the purpose of mass growth for $Z''$.  However, its presence
introduces a non-zero anomaly in $B-L$ and $I_R$. The anomaly free
status of $I_R$ and $B-L$ can be regained by introducing a fourth
flavor $N^c$. With this extension, the dimension 5 operator $(N^c
H'')^2$ is permitted. This gives rise to a Majorana mass contrbibution
$\propto {v''}^2/M_s$ and to a pseudo-Dirac neutrino mass
matrix~\cite{Petcov:1982ya,Wolfenstein:1981kw}. Present limits on
pseudo-Dirac splittings arise from the solar and atmospheric neutrino
measurements. Splitting of less than about $10^{-12} {\rm eV}^2$ (for
$\nu_1$ and $\nu_2$) have no effect on the solar neutrino flux, while
a pseudo-Dirac splitting of $\nu_3$ could be as large as $10^{-4}~{\rm
  eV}^2$ before affecting the atmospheric
neutrinos~\cite{Beacom:2003eu}.  An even stronger bound emerges if we
require the extra relativistic degrees of freedom not to exceed 1 as
indicated by recent cosmological observations. To see this, we note
that the effective thermalization of the right handed neutrinos can
occur through mixing. This will occur if the oscillation length is
less than horizon size during the CMB era. For a typical neutrino mass
of $\sim 0.1~{\rm eV}$, this requires that the Majorana mass is less
than ${\cal O}(10^{-25}~{\rm eV}$).  At present we have no
understanding of the origin of such a hierarchy ({\em i.e.} $10^{-13}$
beyond the ordinary suppression of the Yukawa), and as a consequence
we discard the assignment $I_R= L=+1$ on phenomenological grounds.

Like other broad frameworks for model-building, supersymmetric D-brane
models do not lead uniquely to a single theory.\footnote{Some
  phenomenological aspects of the $U (3)_B \times Sp (1)_L \times
  U (1)_{I_R} \times U (1)_L$ SUSY extension have been discussed
  in~\cite{Kane:2004hm}.}  However, the conjectured models are rather
rigidly constrained, and lead to LHC
predictions that are qualitatively different from the conventional
minimal supersymmetric SM extensions~\cite{Corcella:2012dw}.

We turn now to discuss some specifics of the SUSY extension to our
analysis. The first and obvious change is the modification of the
$\beta$ functions for the running of the couplings. However, these
changes will be minor: the phenomenological requirements at the TeV
scale will effectively fix the $U(1)$ couplings at that scale. Since
unification is not a requirement of D-brane models, the coupling
constants at the string scale will differ somewhat due to the change
in the $\beta$ functions, but string scale couplings do not alter our
phenomenological predictions. The only caveat is to ensure that, as a
result of the enhanced $\beta$ functions, none of the couplings which
comply with TeV data acquire non-pertubative components at the string
scale. We have verified that the variation of the $g'_1(M_s)$
parameter space is hardly noticeable. This gives scarcely any change
in the production cross section and/or branching fractions, even in
the extreme cases shown in Figs.~\ref{fig:CMS} and \ref{fig:ATLAS}, in
which $Z''$ is mostly diagonal in $B-L$ or mostly diagonal in
$I_R$. Furthermore, the milli-weak interactions required to explain the
extra relativistic degrees of freedom during BBN and CMB epochs are
largely independent of these changes.

Much more serious considerations come to light in transcribing the low
energy effective theory into a broken SUSY background. The technical
problem arises most prominently in finding a broken SUSY framework
that will accommodate the hierarchy between the mass of the $Z$ and
the mass of $Z''$. {\it Breaking of the extra $U(1)$ via the Higgs
  mechanism modeled on the radiative breaking of $SU(2) \times U(1)$ driven
  by a large top Yukawa coupling is not an option in the present
  model.} The introduction of an added D-term, a Fayet-Iliopoulos
term, and an extended set of soft breaking masses, requires a sizable
enlargement of the parameter space of the model. In order to
incorporate this parameter space in a phenomenological study it is
imperative to have additional experimental constraints on the SUSY
spectrum.

The approach we have taken here can be regarded as an effective theory
with a new and novel phenomenology, as well as interesting theoretical
characteristics ({\it e.g.}, conservation of $B$ to prevent proton
decay and violation of $L$ without Majorana masses). Of course such an
effective theory requires a high level of fine tuning, which could be
resolved in a more complete broken SUSY framework.  However, we do not
expect the phenomenology to differ in any substantial degree with the
one presented in this paper.

\section{Conclusions}
\label{SEC-VI}

The main purpose of this paper has been to cast D-brane ideology in as
bottoms-up, phenomenologically driven a way as possible.  The energy
scale associated with string physics is assumed to be near the Planck
mass. To develop our program in the simplest way, we considered a
minimal model with gauge-extended sector $U (3)_B \times Sp (1)_L
\times U (1)_{I_R} \times U (1)_L$. The resulting $U (1)$ content
gauges the baryon number $B$, the lepton number $L$, and a third
additional abelian charge $I_R$ which acts as the third isospin
component of an $SU(2)_R$.  Rotation of the $U(1)$ gauge fields to a
basis exactly diagonal in hypercharge $Y$ and very nearly diagonal in
(anomalous) $B$ and (non-anomalous) $I_R$ fixes all mixing angles and
gauge couplings. The anomalous $Z'$ gauge boson obtains a string scale
St\"uckelberg mass via a 4D version of the Green-Schwarz mechanism,
${\rm TeV} \ll M_{Z'} \alt M_s \alt M_{\rm Pl}$. To keep the
realization of the Higgs mechanism minimal, we add an extra $SU(2)$
singlet complex scalar, which acquires a VEV and gives a TeV-scale
mass to the non-anomalous gauge boson $Z''$. It is noteworthy 
that there are no dimension 4 operators involving $H''$ that
contribute to the Yukawa Lagrangian in our D-brane construct. This is
very important since $H''$ carries the quantum numbers of right-handed
neutrino and its VEV breaks lepton number. However, this breaking can
affect only higher-dimensional operators which are suppressed by the
high string scale, and thus there is no phenomenological problem with
experimental constraints for $M_s$ higher than $\sim 10^{14}~{\rm
  GeV}$. Since all freedom of determining coupling constant and mixing angles
has been exercised, there remains only constraints on the possible
value of $M_{Z''}$. We have shown  that $M_{Z''} \approx 3 - 4~{\rm TeV}$
saturates current limits from the CMS  and
ATLAS  collaborations. We have also shown that
for $M_{Z''} \alt 5~{\rm TeV}$, LHC14  will reach
discovery sensitivity $\agt 5\sigma$.

Armed with our D-brane construct, we developed a dynamic explanation
of recent hints that the relativistic component of the energy during
the CMB and BBN epochs is equivalent to about 1 extra Weyl
neutrino. Requiring that the $B-L$ current be anomaly free implies
existence of 3 right-handed Weyl neutrinos. The task then reverts to
explain why there are not 3 additional r.d.o.f.  We showed that for
certain ranges of $M_{Z''}$ the decoupling of the $\nu_R$'s occurs
during the course of the quark-hadron crossover transition, just so
that they are only partially reheated compared to the $\nu_L$'s ---
the desired outcome.  Roughly speaking, if decoupling requires a
freezout of the annihilation channel (loss of chemical equilibrium),
then for a $Z''$ which is mostly $I_R$, $3.6~{\rm TeV} < M_{Z''}<
4.8~{\rm TeV}$, whereas for a $Z''$ which is mostly $B-L$, $4.5~{\rm
  TeV} < M_{Z''}< 6.1~{\rm TeV}$. This range will be probed at LHC14.
If thermal equilibrium via scattering is sufficient, for a $Z''$ which
is mostly $I_R$, $5.4~{\rm TeV} < M_{Z''}< 7.4~{\rm TeV}$, and for a
$Z''$ which is mostly $B-L$, $6.3~{\rm TeV} < M_{Z''}< 8.2~{\rm
  TeV}$. To carry out this program, we needed to make use of some high
statistics lattice simulations of a QCD plasma in the hot phase,
especially the behavior of the entropy during the
confinement-deconfinement changeover.  Interestingly, the behavior of
the trace anomaly (shown in Fig.~15 of~\cite{Bazavov:2009zn}), which
is very sensitive to the nature of the crossover region, shows a sharp
peak at $200~{\rm MeV}$ and our range for $T_{\rm dec}$ straddles this
region.

Throughout this paper we remained agnostic with respect to SUSY
breaking and the  details of the low energy effective
potential. However, we do subject the choice of quantun numbers for
$H''$ to the stringent holonomic constraints of the superpotential at
the string scale.  This forbids the simultaneous presence of scalar
fields and their complex conjugate. As an illustration, if the quantum
numbers of $H''$ are those of $N_R^c$, then higher dimensional
operators such as $\overline N_R N_R^c {H''}^2$, which can potentially
generate a Majorana mass, are absent. Because of holonomy this absence
cannot be circunvented by including $\overline N_R N_R^c
{H''}^{*2}$. 

In summary, we have studied the $U(1)$ phenomenology of D-brane models
endowed with a high mass string scale. We have incorporated some
elements of SUSY, discussing evolution of the gauge couplings to the
string scale and enforcing the holonomic constraints on the
superpotential. We have shown that LHC8 data set upper limits on the
mass of the $Z''$ gauge boson: $M_{Z''} \alt 3 - 4~{\rm TeV}$. We have
also shown that $Z''$ milli-weak interactions, which are within reach
of LHC14, could play an important role in observational cosmology. It
is important to stress that the $Z''$ production cross section and its
branching fractions are {\it universal} and have been evaluated in a
parameter-free manner. Therefore, the $U(1)$ phenomenology presented
in this paper is completely independent of the details of the
compactification scheme, such as the configuration of branes, the
geometry of the extra dimensions, and whether the low energy theory is
supersymmetric or not.

\section*{Acknowledgments}
L.A.A.\ is supported by the U.S. National Science Foundation (NSF)
under CAREER Grant PHY-1053663. I.A.\ is supported in part by the
European Commission under the ERC Advanced Grant 226371 and the
contract PITN-GA-2009- 237920. H.G.\ and T.R.T.\ are supported by NSF
Grant PHY-0757959.  X.H.\ is supported in part by the National
Research Foundation of Korea grants 2005-009-3843, 2009-008-0372, and
2010-220-C00003. D.L.\ is partially supported by the Cluster of
Excellence "Origin and Structure of the Universe", in Munich. D.L.\
and T.R.T.\ thank the Theory Department of CERN for its hospitality.
L.A.A.\ and H.G.\ thank the Galileo Galilei Institute for Theoretical
Physics for the hospitality and the INFN for partial support during
the completion of this work.  Any opinions, findings, and conclusions
or recommendations expressed in this material are those of the authors
and do not necessarily reflect the views of the National Science
Foundation.

\end{document}